\documentclass[draftcls,onecolumn]{IEEEtran}
\usepackage{amsmath,epsfig,amssymb,dsfont,cite}
\usepackage[]{algorithm2e}
\usepackage{float}
\usepackage{subfigure}

\newcommand{\RR}{{\rm I\!R}}

\newtheorem{proposition}{Proposition}
\newtheorem{remark}{Remark}
\newtheorem{theorem}{Theorem}

\ifCLASSINFOpdf
\else
\fi

\begin{document}
%
\title{Independent Resampling Sequential Monte Carlo Algorithms}
%
%
%

\author{Roland Lamberti, Yohan Petetin, Fran\c cois Desbouvries, and Fran\c cois Septier 
\thanks{R. Lamberti, Y. Petetin and F. Desbouvries are with 
Samovar, Telecom Sudparis, CNRS, Universit\'e Paris-Saclay, 9 rue Charles Fourier, 91011 Evry, France}
\thanks{F. Septier is with Telecom Lille \& UMR CNRS CRIStAL, Rue Guglielmo Marconi, 59653 Villeneuve d'Ascq, France}
}

\maketitle

\begin{abstract}
Sequential Monte Carlo algorithms, or Particle Filters, 
are Bayesian filtering algorithms which propagate in time a discrete 
and random approximation of the a posteriori distribution of interest.
Such algorithms are based on Importance Sampling with a bootstrap
resampling step which aims at struggling against weights degeneracy.
However, in some situations (informative measurements, high dimensional model),
the resampling step can prove inefficient. 
In this paper, we revisit the fundamental resampling mechanism which leads us back to Rubin's static 
resampling mechanism. 
We propose an alternative rejuvenation scheme in which the resampled 
particles share the same marginal distribution as in the classical setup,
but are now independent. 
This set of independent particles provides a new alternative to compute a moment 
of the target distribution and the resulting estimate is analyzed through a CLT.
We next adapt our results to the dynamic case and propose a particle filtering
algorithm based on independent resampling.
This algorithm can be seen as a particular auxiliary particle filter algorithm with
a relevant choice of the first-stage weights and instrumental distributions. 
Finally we validate our results via simulations which carefully 
take into account the computational budget.

\end{abstract}


%
\IEEEpeerreviewmaketitle

\section{Introduction}
%
%
%
%
\IEEEPARstart{L}{et} 
$\{X_k \in \mathbb{R}^m  \}_{k \geq 0}$ (resp. 
$\{Y_k \in \mathbb{R}^n  \}_{k \geq 0}$) be a hidden (resp. observed) process.
Let $X_{0:k}$, say, denote $\{ X_i, 0 \leq i \leq k \}$,
$x_{0:k} = \{ x_i, 0 \leq i \leq k \}$,
and let $p(x)$ (resp. $p(x|y)$), say, 
denote the probability density function (pdf) of random variable (r.v.) $X$ (resp. of $X$ given $Y=y$);
capital letters are used for r.v. and lower case ones for their realizations.
We assume that $\{ (X_k, Y_k) \}_{k \geq 0}$ is a Hidden Markov chain, i.e. that
\begin{eqnarray}
\label{HMC}
p(x_{0:k},y_{0:k})=p(x_0)\prod_{i=1}^k f_{i}(x_i|x_{i-1}) \prod_{i=0}^kg_i(y_i|x_i)\text{.}
\end{eqnarray}Roughly speaking,
pdf $f_{k}(x_k|x_{k-1})$ describes the dynamical evolution
of the Markovian hidden process $\{X_k \}_{k \geq 0}$ between time $k-1$
and time $k$ while the likelihood $g_k(y_k|x_k)$ describes 
the relation at time $k$ between an observation $y_k$ 
and the associated hidden state $x_k$.
We address the problem of computing 
a moment of some function $f({\bf .})$ w.r.t. the filtering
pdf $p(x_k|y_{0:k})$, 
i.e. the pdf of the hidden state given the past observations:
\begin{equation}
\label{Theta-HMC}
\Theta_k = \int f(x_k) p(x_k|y_{0:k}){\rm d}x_k \text{.}
\end{equation}

As is well known,
$\Theta_k$ can be exactly computed
only in very specific models,
and one needs to resort to approximations in the general case.
In this paper, 
we focus on a popular class of approximations
called sequential Monte Carlo (SMC) algorithms
or Particle Filters (PF),
see e.g.
\cite{gordon-salmond-smith,livredoucet,arulampalam} 
PF propagate over time a set of $N$ Monte Carlo (MC)
weighted samples $\{w_{k}^i,x_k^i\}_{i=1}^N$ which defines 
a discrete approximation 
$\sum_{i=1}^N w_{k}^i \delta_{x_k^i}$
of $p(x_k|y_{0:k})$
and enables to compute an estimate $\widehat{\Theta}_k$ of $\Theta_k$:
\begin{equation}
\label{Theta-estimate}
\widehat{\Theta}_k = \sum_{i=1}^N w_k^if(x_k^i) \text{.}
\end{equation}
More precisely, the computation of the set $\{w_{k}^i,x_k^i\}_{i=1}^N$
is based on the sequential application of 
the Importance Sampling (IS) mechanism \cite{Gewecke}.
This mechanism consists in sampling particles according to an importance distribution 
and next weighting these samples 
in order to correct the discrepancy between
the target and the importance distribution. 
However 
the direct sequential
application of the IS mechanism in model \eqref{HMC} 
fails in practice
since after a few time steps most weights get close to to zero,
while only a few particles have non neglictible weights.
Consequently IS alone becomes more and more inefficient 
since a lot of computational effort is devoted to sampling particles 
which will hardly contribute to the estimate $\widehat{\Theta}_k$ in \eqref{Theta-estimate}.

As is well known, 
a traditional rescue against weights degeneracy
consists in {\sl resampling} the particles
(- either at each time step
or depending on some criterion such as the number of efficient particles
\cite{kong1994} 
\cite{liu-chen1995}
\cite{liu1996metropolized}
\cite{cornebise}),
i.e. of re-drawing each particle with a probability equal to its weight.
This yields the class of
Sampling Importance Resampling (SIR) algorithms
\cite{smith-gelfand} \cite{gordon-salmond-smith}
\cite{douc-cappe-resampling} \cite{Li:2015fl}.
This resampling (i.e., bootstrap) mechanism has proved to be beneficial in the long run,
but its instantaneous effects are mitigated;
though the resampling step 
indeed discards particles with low weights
(such particles are likely never to be resampled),
particles with significant weights are resampled several times,
which results in dependency among the resampled points and support shrinkage.
Consequently, particle filters based on the resampling mechanism can give
poor results in some Markovian models \eqref{HMC}, such as informative
models where the likelihood $g_k(y_k|x_k)$ is sharp.
Our aim in this paper is thus to revisit this key rejuvenation scheme 
in order to design new PF algorithms 
which would keep the benefits of the resampling mechanism,
while avoiding the local impoverishment 
of the resulting MC approximation of the filtering distribution.

To that end we begin 
with revisiting the SIR mechanism at one single time step $k \rightarrow k+1$.
This leads us back to an analysis of Rubin's static SIR mechanism
\cite[\S 2]{rubin1988}
\cite{gelfand-smith}
\cite{smith-gelfand}
\cite[\S 9.2]{Cappeetal},
in which,
roughly speaking,
one obtains samples $x^j$ 
approximately drawn from a target distribution $p$
by drawing intermediate samples $\{ \tilde{x}^i \}_{i=1}^N$ from an instrumental distribution $q$,
and next selecting $x^j$ among $\{ \tilde{x}^i \}_{i=1}^N$ with a probability proportional to $\frac{p(\tilde{x}^i)}{q(\tilde{x}^i)}$.
We first observe 
that the samples $\{ x^j \}$ produced by this SIR mechanism
are dependent and marginally distributed from some compound pdf 
$\tilde{q}_N = \phi(p, q, N)$ which takes into account the effects of both pdfs $p$ and $q$.
Here the dependency is detrimental,
because samples that would be i.i.d from $\tilde{q}_N$ 
would produce,
whichever the number of sampled and resampled particles,
a moment estimate with reduced variance;
this result 
is further illustrated by a central limit theorem (CLT)
which is compared to the existing CLTs 
for the static IS estimate
(based on the pre-resampling samples $\{ \tilde{x}^i \}_{i=1}^N$),
on the one hand,
and for the SIR estimate 
(based on the post-resampling ones $\{ x^j \}_{j=1}^{M_N}$),
on the other hand.

We next propose a procedure to obtain i.i.d. samples from $\tilde{q}_N$,
which leads to the computation of two point estimates of $\Theta= \int f(x) p(x) {\rm d} x$.
The first one is based on unweighted i.i.d. samples 
and is an improved version of the classical (i.e., dependent) SIR estimate; 
the second one is based on post-resampling-weighted i.i.d. samples
and can be seen as new IS estimate,
based on the compound pdf $\tilde{q}_N$.
Finally we adapt these results 
to the sequential computation of $\Theta_k$ 
in model \eqref{HMC}. 
We thus propose two new PF algorithms.
One of them has an interesting interpretation in
terms of Auxiliary Particle Filter (APF);
more precisely,
that algorithm naturally produces a relevant importance mixture distribution 
from which it is easy to sample.
We finally illustrate our results via simulations, and carefully compare our algorithms 
with existing ones in terms of Root Mean Square Error (RMSE) 
and computational cost.
The rest of this paper is organized as follows. 
Section \ref{section-1} is devoted to the static case.
In section \ref{section-2} we address the sequential case, 
and derive new PF based 
on the results of section \ref{section-1}.
In section \ref{section-3} we perform simulations and
discuss implementation issues,
and we end the paper with a conclusion.

\section{IS with resampling viewed as a compound IS scheme}
\label{section-1}


As recalled in the introduction, 
resampling from time to time 
is a standard rescue when applying IS in the sequential case.
In this section we thus focus on one such time step $k \rightarrow k+1$.
This amounts to revisiting Rubin's static SIR mechanism (see section \eqref{review-SIR}),
which consists in resampling points $\{ x_i\}_{i=1}^{M_N}$ 
from the weighted distribution
$\sum_{i=1}^N w_i \delta_{\tilde{x}_i}$
where 
$\tilde{x}_i \stackrel{i.i.d.}{\sim} q$
and the pre-resampling weights 
$w_i \propto \frac{p(\tilde{x}_i)}{q(\tilde{x}_i)}$ with $\sum_{i=1}^N w_i = 1$.
As is well known, 
when $N \rightarrow \infty$ 
the resampled points $\{ x_i\}_{i=1}^{M_N}$ become asymptotically i.i.d. from the target distribution $p$.
For finite $N$ however, 
these samples are dependent and drawn from some pdf $\tilde{q}_N$ 
which differs from $p$ and can indeed be seen as a compound IS density $\tilde{q}_N = \phi(p, q, N)$ 
produced by the succession of the sampling ({\sl S}), weighting ({\sl W}) and resampling ({\sl R}) steps.
We discuss on the benefits of drawing independent samples from $\tilde{q}_N$ (see section \ref{independent-SIR}),
and next on reweighting these independent samples 
with post-resampling weights $w'_i \propto \frac{p(x_i)}{\tilde{q}_N(x_i)}$
(see section \ref{reweighting}).
In all this section we assume the scalar case for simplicity.
We end the section with a summary (see section \ref{summary-static}).

\subsection{The dependent SIR mechanism}
\label{review-SIR}

Let us begin with a brief review of Rubin's classical SIR sampling mechanism
and of the properties of the sampled and resampled particles.

\subsubsection{Properties of the sampled particles $\{ \tilde{x}^i \}_{i=1}^{N}$}

In the context of this paper we first recall the principle of IS.
Let $p(x)$ be a probability density function and
assume that we want to compute 
\begin{equation}
\label{Theta}
\Theta= \int f(x) p(x) {\rm d} x = {\rm E}_p(f(X)) \text{.}
\end{equation}
In the Bayesian framework $p(x)$ is generally only known 
up to a constant, i.e. $p(x) \propto p_u(x)$ (subscript $u$ is for unnormalized)
and it is not possible to obtain samples directly drawn from $p(x)$.
A solution is to introduce an importance distribution $q(x)$ which 
satisfies $q(x)>0$ when $p(x)>0$ and to rewrite $\Theta$
as the ratio of two expectations w.r.t. $q$,
\begin{equation}
\Theta= \frac{\int f(x) \frac{p_u(x)}{q(x)} q(x) {\rm d} x}{\int \frac{p_u(x)}{q(x)} q(x){\rm d} x} = 
\frac{{\rm E}_q(f(X)\frac{p(X)}{q(X)})}{{\rm E}_q(\frac{p(X)}{q(X)})} \text{.}
\end{equation}
Next, each expectation is approximated by a Monte Carlo method based on $N$ i.i.d.
samples $(\tilde{x}^1,\cdots,\tilde{x}^N)$ drawn from $q(.)$;
the IS estimate of $\Theta$ is given by
\begin{equation}
\label{theta-is}
\widehat{\Theta}_N^{{\rm IS}}= \sum_{i=1}^N w^i f(\tilde{x}^i) 
= {\rm E}_{\hat{p}}(f(X)) 
\end{equation}
where 
\begin{equation}
\label{p_hat}
\hat{p}(x) = \sum_{i=1}^N w^i \delta_{\tilde{x}^i}(x)
\end{equation}
and where
$w^i$ (the $i$-th normalized importance weight) reads
\begin{equation}
w^i= \frac{\frac{p_u(\tilde{x}^i)}{q(\tilde{x}^i)}}{\sum_{j=1}^N \frac{p_u(\tilde{x}^j)}{q(\tilde{x}^j)}} = \frac{\frac{p(\tilde{x}^i)}{q(\tilde{x}^i)}}{\sum_{j=1}^N \frac{p(\tilde{x}^j)}{q(\tilde{x}^j)}} \text{.}
\end{equation}
As is well known \cite{Gewecke}, 
under mild assumptions
\begin{equation}
\label{LLN-IS}
\widehat{\Theta}_N^{{\rm IS}} \stackrel{\rm a.s.}{\rightarrow} \Theta,
\end{equation} 
and a CLT 
is available too (${\mathcal{D}} {\rightarrow}$ denotes the convergence in distribution):
\begin{equation}
\label{TCL-IS}
\sqrt{N} (\widehat{\Theta}^{\rm IS}_N-\Theta)  \overset {\mathcal{D}} {\rightarrow}  \mathcal{N} \left(0,{\rm E}_q \left(\frac{p^2(X)}{q^2(X)}(f(X)-\Theta)^2\right) \right).
\end{equation}

\subsubsection{Properties of the resampled particles $\{ x^i \}_{i=1}^{M_N}$}

From 
\eqref{LLN-IS} and
\eqref{TCL-IS},
$\hat{p}$ can be seen as a discrete approximation of the target density $p$, 
and one expects that for large $N$, (re)sampling from $\hat{p}$ would produce samples approximately drawn from $p$.
This is the rationale of Rubin's SIR mechanism
\cite[\S 2]{rubin1988},
\cite{gelfand-smith},
\cite{smith-gelfand},
\cite[\S 9.2]{Cappeetal}.
More precisely, 
let us as above draw $N$ i.i.d. samples $\tilde{x}^i$ from $q$,
and next $M_N$ i.i.d samples $x^i$ from $\hat{p}$ in \eqref{p_hat}.
It is indeed well-known (see \cite{smith-gelfand} \cite{rubin1988}) that when $N \rightarrow \infty$,
each r.v. $x^i$ produced by this mechanism converges in distribution to $p(.)$,
so Rubin's technique can be seen as a two-step sampling mechanism 
which transforms samples drawn from $q$ into samples (approximately) drawn from $p$.

This convergence result can be completed by a CLT
which involves the estimate of $\Theta$ based on the unweighted set $\{(\frac{1}{M_N},x^i)\}_{i=1}^{M_N}$:
\begin{equation}
\label{theta-sir}
\widehat{\Theta}^{\rm SIR}_{M_N}=\frac{1}{M_N} \sum_{i=1}^{M_N} f(x^i)\text{.}
\end{equation}
Let
$N \rightarrow \infty$, let $M_N$ be a non decreasing sequence with
$M_N \rightarrow \infty$, and let
${\rm lim}\frac{N}{M_N} = \alpha > 0$ (possibly $\infty$);
then under mild conditions
(see e.g. 
\cite[\S 9]{Cappeetal})
\begin{align}
\label{TCL-SIR}
\sqrt{M_N}(\widehat{\Theta}^{\rm SIR}_{M_N}-\Theta) & \overset {\mathcal{D}} {\rightarrow}  \mathcal{N}(0,{\rm var}_p(f(X))+
 \alpha^{-1}{\rm E}_q\left((\frac{p^2(X)}{q^2(X)}(f(X)-\Theta)^2)\right)) \text{.} 
\end{align}
If $\alpha \rightarrow \infty$ then the asymptotic variance tends to ${\rm var}_p(f(X))$, 
which shows that the
SIR estimate asymptotically has the same behavior as a crude Monte Carlo estimate directly
deduced from $M_N$ samples according to the target distribution $p(.)$,
provided the number $N$ of intermediate samples is large compared to $M_N$.

However, for computational reasons, 
the number of samples $N$ and $M_N$ should
not be too large in practice. 
Consequently we now focus on the samples produced by the SIR 
procedure from a non asymptotical point of view and
we have the following result (the proof is given in the Appendix).

\begin{proposition}
\label{prop-q-tilde-static}
Let us consider the samples $\{x^i\}_{i=1}^{M_N}$ produced by 
the SIR mechanism described above. Then these samples
are identically  distributed according to a pdf 
$\tilde{q}_{N}$, with
\begin{eqnarray}
\label{q-tilde-static}
\tilde{q}_N(x) & = & N h_N(x) q(x) \text{,}  \\
\label{h-static}
h_N(x)         & = & 
\int \int \frac{\frac{p(x)}{q(x)}}{\frac{p(x)}{q(x)} + \sum_{l=1}^{N-1} \frac{p(x^l)}{q(x^l)}} \prod_{l=1}^{N-1} q(x^l) {\rm d}x^l \text{.}
\end{eqnarray}
\end{proposition}

So for fixed sample size $N$,
the SIR mechanism produces dependent samples $\{ x^i \}_{i=1}^{M_N}$ distributed from $\tilde{q}_{N}$
(these samples are independent given the intermediate set $\{ \tilde{x}^i \}_{i=1}^{N}$,
but become dependent when this conditioning is removed).
In practice, this dependency results in support shrinkage since, by construction,
an intermediate sample $\tilde{x}^i$ can be resampled several times,
and $\{ x^i \}_{i=1}^{M_N}$ is a subset of $\{ \tilde{x}^i \}_{i=1}^{N}$.
For instance let $M_N=N$. 
If we assume that $w^j=1$ for some $j$ and $w^i=0$  for $i \neq j$,
then $x^i=\tilde{x}^j$ for all $i$. 
By contrast, if $w^i=1/N$ for all $i$, 
then the average number of different samples $\{x^i\}_{i=1}^N$ is approximately $N/3$ \cite{petetin-sco}.
Nevertheless the resampling step remains useful in a dynamic setup (see section \ref{section-2}): 
even though locally it leads to an impoverishment of the diversity, 
this step is critical for recreating diversity at the next time step.

\subsection{The independent SIR mechanism}
\label{independent-SIR}
%
%
%
%

Observe that the two factors in \eqref{q-tilde-static}
reflect the effects of the sampling and resampling step:
pdf $q$ is used in the {\sl S} step,
while $h_N(x)$, 
which can be interpreted as the conditional expectation 
of a normalized importance weight when its associated particle is $x$,
results from the ({\sl W},{\sl R}) steps.
So particles drawn from $\tilde{q}_N$ are likely to be in regions where 
1) $q$ is large (since these particles have first been sampled); and 
2) which have also been resampled because their associated weight was large enough.
Now our objective is to propose an alternative mechanism which,
in the sequential case, 
will produce the same positive effect as the classical SIR mechanism
(i.e. fighting against weight degeneracy by eliminating the samples 
with weak importance weights),
while ensuring the diversity of the final support.
Such a support diversity is ensured if we draw samples {\sl independently} from the continuous pdf $\tilde{q}_N(.)$.
We first study the potential benefits of this sampling mechanism (see section \ref{stat-properties})
and next discuss its implementation (see section \ref{sampling-independent}).

\subsubsection{Statistical properties}
\label{stat-properties}
Let us now assume that we have at our disposal 
a set of $M_N$ i.i.d. samples $\{\overline{x}^i\}_{i=1}^{M_N}$ drawn
from $\tilde{q}_N(.)$ defined in \eqref{q-tilde-static} \eqref{h-static}. 
Before addressing the practical computation of such a set (see section \ref{sampling-independent}), 
let us study its properties by considering 
the crude estimate of $\Theta$ based on these $M_N$ i.i.d samples:
\begin{equation}
\label{theta-isir}
\widehat{\Theta}^{\rm I-SIR}_{M_N}=\frac{1}{M_N} \sum_{i=1}^{M_N} f(\overline{X}^i)\text{.}
\end{equation} 
(I in notation I-SIR stands for independent).
Our aim is to compare 
$\widehat{\Theta}^{\rm I-SIR}_{M_N}$ to
$\widehat{\Theta}^{\rm SIR}_{M_N}$, 
and more generally
$\widehat{\Theta}^{\rm IS}_{N}$,
$\widehat{\Theta}^{\rm SIR}_{M_N}$ and
$\widehat{\Theta}^{\rm I-SIR}_{M_N}$.
We first have the following result
(the proof is given in the Appendix).

\begin{proposition}
\label{prop-static-2}
Let us consider the three estimates 
$\widehat{\Theta}^{\rm IS}_{N}$,
$\widehat{\Theta}^{\rm SIR}_{M_N}$ and
$\widehat{\Theta}^{\rm I-SIR}_{M_N}$
defined in \eqref{theta-is}, \eqref{theta-sir} and \eqref{theta-isir} 
respectively.
Then 
\begin{eqnarray}
\label{egal-esperances}
{\rm E}(\widehat{\Theta}^{\rm IS}_{N}) &=& {\rm E}(\widehat{\Theta}^{\rm SIR}_{M_N})={\rm E}(\widehat{\Theta}^{\rm I-SIR}_{M_N}) \text{,}
\\
\label{dependent-vs-independent}
{\rm var}(\widehat{\Theta}^{\rm SIR}_{M_N}) &=& {\rm var}(\widehat{\Theta}^{\rm I-SIR}_{M_N}) + \frac{M_N-1}{M_N}{\rm var}(\widehat{\Theta}^{\rm IS}_{N}) \text{.}
\end{eqnarray}
\end{proposition}

%
Equation \eqref{dependent-vs-independent} ensures that an estimate based on independent samples obtained
from $\tilde{q}_N$ outperforms the classical SIR estimate; the gain 
of $\widehat{\Theta}^{\rm I-SIR}_{M_N}$ w.r.t. $\widehat{\Theta}^{\rm SIR}_{M_N}$
depends on the variance of ${\rm var}(\widehat{\Theta}^{\rm IS}_{N})$.
On the other hand it is well known (see e.g. \cite[p. 213]{Cappeetal}) that
${\rm var}(\widehat{\Theta}^{\rm SIR}_{M_N}) =$
${ \rm var}(\widehat{\Theta}^{\rm IS}_N)+ {\rm E}({\rm var}(\widehat{\Theta}^{\rm SIR}_{M_N}|\{\tilde{x}^i\}_{i=1}^N))$;
so both 
$\widehat{\Theta}^{\rm I-SIR}_{M_N}$ and
$\widehat{\Theta}^{\rm IS}_{N}$
are preferable to 
$\widehat{\Theta}^{\rm SIR}_{M_N}$.

On the other hand,
comparing the variance of $\widehat{\Theta}^{\rm IS}_{N}$
to that of $\widehat{\Theta}^{\rm I-SIR}_{M_N}$ is more difficult, 
because we have to compare 
$\frac{1}{M_N} {\rm var}_{\tilde{q}_N}(f(\overline{X}))$ to
${\rm var}(\sum_{i=1}^N w^i(\tilde{X}^1,\cdots,\tilde{X}^N)f(\tilde{X}_i))$ 
where $\tilde{X}^i \overset{\rm i.i.d} {\sim} q(.)$. 
However, we have
the following CLT (the proof is given in the Appendix).
\begin{theorem}
\label{prop-tcl}
Let us consider the independent SIR estimate
defined in \eqref{theta-isir}.
Let assume that 
$N   \rightarrow \infty$, 
$M_N$ is a non decreasing sequence with
$M_N \rightarrow \infty$
and $\displaystyle{\lim_{N \rightarrow \infty }}\frac{N}{M_N} = \alpha > 0$.
Then $\widehat{\Theta}^{\rm I-SIR}_{M_N}$ satisfies 
\begin{equation}
\label{TCL-1}
\sqrt{M_N}(\widehat{\Theta}^{\rm I-SIR}_{M_N}- \Theta) \overset {\mathcal{D}} {\rightarrow}  \mathcal{N} \left(0,{\rm var}_p (f(X))\right) \text{.}
\end{equation}
\end{theorem}

Let us comment this result.
First Theorem \ref{prop-tcl}
enables again to compare 
$\widehat{\Theta}^{\rm I-SIR}_{M_N}$ to
$\widehat{\Theta}^{\rm SIR}_{M_N}$.
Comparing 
\eqref{TCL-SIR} and \eqref{TCL-1}
confirms \eqref{dependent-vs-independent},
since the asymptotic variance of 
$\widehat{\Theta}^{\rm I-SIR}_{M_N}$ is always lower than that of 
$\widehat{\Theta}^{\rm SIR}_{M_N}$.
Also note that in the independent case 
the asymptotic variance of 
$\widehat{\Theta}^{\rm I-SIR}_{M_N}$ no
longer depends on $\alpha > 0$.

Next Theorem \ref{prop-tcl}
also gives some elements for comparing
$\widehat{\Theta}^{\rm I-SIR}_{M_N}$ to
$\widehat{\Theta}^{\rm IS}_{N}$. 
Let for simplicity $M_N = N$. 
Then the comparison of both estimates relies on that of the asymptotic variances in 
\eqref{TCL-IS} and 
\eqref{TCL-1}:
\begin{eqnarray}
\label{var-asympt-IS}
\sigma^{2, \rm IS}_{\infty} (q) & = &
{\rm E}_q \left(\frac{p^2(X)}{q^2(X)}(f(X)-\Theta)^2\right), \\
\label{var-asympt-ISIR}
\sigma^{2, \rm I-SIR}_{\infty} & = & {\rm var}_p (f(X)).
\end{eqnarray} 
For a given target pdf $p({\bf .})$ and function $f({\bf .})$,
$\sigma^{2, \rm IS}_{\infty}(q)$ depends on the importance pdf $q({\bf .})$
and is well known \cite[\S 2.9]{these-hesterberg}
\cite[Theorem 3]{Gewecke}
to be minimum for $q^{\star}(x) \propto p(x)|f(x) - \Theta|$; 
for that $q^{\star}$,
$\sigma^{2, {\rm IS}}_{\infty}(q^{\star}) =  (\int |f(x)-\Theta|p(x){\rm d}x)^2 \leq {\rm var}_p (f(X))$,
so
$\widehat{\Theta}^{\rm IS}_{N}(q^{\star})$ outperforms 
$\widehat{\Theta}^{\rm I - SIR}_{N}$ for large values of $N$.
On the other hand for other importance distributions
$\sigma^{2, \rm IS}_{\infty}(q)$ may become larger than 
$\sigma^{2, \rm I - SIR}_{\infty}$. 
Also note that the variances in 
\eqref{var-asympt-IS} and
\eqref{var-asympt-ISIR}
depend on function $f({\bf .})$;
on the other hand, for large $N$,
$\widehat{\Theta}^{\rm I-SIR}_{N}$ has the same behavior as a crude estimate built from samples drawn from $p(.)$ 
and so is adapted for a large class of functions $f({\bf .})$.

\subsubsection{Sampling procedure}
\label{sampling-independent}

It remains to describe a procedure to obtain i.i.d. samples 
from $\tilde{q}_N$. 
Algorithm 
\ref{algo-ISIR} ensures that the final 
samples $\{x^i,\cdots,x^{M_N}\}$ are drawn independently from 
$\tilde{q}_N$.

\begin{algorithm}
\SetKwInOut{KwInput}{Input}
\KwInput{an importance distribution $q$, $N$ and $M_N$}
\KwResult { $\{x^i\}_{i=1}^{M_N}$ $\overset{\rm i.i.d.} {\sim} \tilde{q}_N$}
\For{$1 \leq i \leq M_N$} 
{
\For{$1 \leq j \leq N$} 
{
 
\textbf{S.} $\tilde{x}^{i,j} \sim q(.)$;  \\
\textbf{W.} $w^{i,j} \propto  p_u(\tilde{x}^{i,j})/q(\tilde{x}^{i,j})$, $\sum_{j=1}^N w^{i,j}=1$;
}
\textbf{R.} $x^i \sim \sum_{j=1}^N w^{i,j} \delta_{\tilde{x}^{i,j}}$
}
\caption{The independent SIR algorithm}
\label{algo-ISIR}
\end{algorithm}



Compared to the classical SIR procedure, the independent 
SIR algorithm described in Algorithm \ref{algo-ISIR} relies on 
a sampling step of $N \times M_N$ intermediate samples
$\tilde{x}$ and $M_N$ independent resampling steps. 
Consequently, for a given budget of sampling and resampling
steps, the independent procedure should be compared with a classical
SIR one in which we sample $N \times M_N$ points and resample $M_N$ of them. 
In this last
case, we obtain $M_N$ dependent samples drawn from
$\tilde{q}_{N \times M_N}$. 
First, 
using \eqref{TCL-SIR} with 
$\alpha=\lim_{N\rightarrow \infty} \frac{N \times M_N}{M_N}=\infty$,
we see that both estimates $\widehat{\Theta}^{\rm I-SIR}_{M_N}$
and $\widehat{\Theta}^{\rm SIR}_{M_N}$ with $N\times M_N$ intermediate
samples have the same asymptotic behavior.
However the independent procedure 
can be easily parallelized 
because the resampling steps are by nature
independent contrary to the SIR procedure where the 
$N \times M_N$ intermediate samples are directly
resampled.

\subsection{Reweighting the independent samples?}
\label{reweighting}

We finally discuss the final weights which are attributed 
to the resampled particles. 
In the SIR procedure, each 
final sample is weighted by $1/M_{N}$. 
From an IS point of view, 
this weighting traduces the fact 
that the final samples become drawn from the target distribution
$p(.)$ and independent when $N \rightarrow \infty$ \cite{rubin1988}.
Moreover the 
convergence results of $\widehat{\Theta}^{\rm I-SIR}_{M_N}$ 
to $\Theta$ (see e.g. \cite{delmoral-livre} \cite{Cappeetal}) confirm
that these weights are valid from an asymptotical point of view.
In the independent SIR procedure, the only difference
is that the final samples are independent, even from a
non-asymptotical point of view.

Now, if $N$ is finite, one can wonder if weights $1/M_N$ are optimal. 
In Algorithm \ref{algo-ISIR}, 
samples $\{X^i\}_{i=1}^{M_N}$ are independent and
sampled from $\tilde{q}_N$. Consequently, for a given $N$, $\tilde{q}_N$
can be seen as a post-resampling 
compound importance distribution $\tilde{q}_N = \phi(p, q, N)$, 
and a final sample $x^i$ should be weighted by a post-resampling weight
proportional 
to $p_u(x^i)/\tilde{q}_N(x^i)$. 
This yields a new
estimate 
$\widehat{\Theta}^{\rm I-SIR-w}$
of \eqref{Theta} (superscript $w$ stands for weighted)
\begin{equation}
\label{theta-ISIR-2}
\widehat{\Theta}^{\rm I-SIR-w}_{M_N} = \sum_{i=1}^{M_N}  \frac{\frac{p_u(x^i)}{\tilde{q}_N(x^i)}}{\sum_{j=1}^{M_N}\frac{p_u(x^j)}{\tilde{q}_N(x^j)}}f(x^i) \text{,}
\end{equation}
which coincides with the IS estimate \eqref{theta-is}
with importance distribution $\tilde{q}_N(.)$.
It is difficult to compare $\widehat{\Theta}^{\rm I-SIR}_{M_N}$ and
$\widehat{\Theta}^{\rm I-SIR-w}_{M_N}$ because the expression of the weights in this last
case depends on $N$. 
However, it is interesting to note that contrary to 
$\widehat{\Theta}^{\rm I-SIR}_{M_N}$, $M_N$ impacts the bias of the estimate $\widehat{\Theta}^{\rm I-SIR-w}_{M_N}$.
For example, if we set $N=1$ (so $q_N=q$) and $M_1$ is arbitrary then $\widehat{\Theta}^{\rm I-SIR-w}_{M_1}$
coincides with the IS estimate with $M_1$ i.i.d. samples drawn from $q$ 
while the unweighted estimate
$\widehat{\Theta}^{\rm I-SIR}_{M_1}$ is a crude estimate of $\int f(x) q(x) {\rm d}x$
and is not adapted for the estimation of $\Theta$.
More generally, 
using the delta method to approximate  ${\rm E}(\widehat{\Theta}^{\rm I-SIR}_{M_N})$
and ${\rm E}(\widehat{\Theta}^{\rm I-SIR-w}_{M_N})$ \cite{Liu-bouquin} we observe that
\begin{align}
\label{biais1}
& {\rm E}(\widehat{\Theta}^{\rm I-SIR}_{M_N}) \! = \! {\rm E}(\widehat{\Theta}^{\rm IS}_{N}) \! \approx \! \Theta  \! - \! \frac{1}{N}{\rm E}_q \left(\frac{p^2(X)}{q^2(X)}(f(x) \! - \! \Theta)\right) \!\! \text{,} \\
\label{biais2}
& {\rm E}(\widehat{\Theta}^{\rm I-SIR-w}_{M_N}) \! \approx  \! \Theta \! - \! \frac{1}{M_N}{\rm E}_{\tilde{q}_N} \left(\frac{p^2(X)}{\tilde{q}_N^2(X)}(f(x)-\Theta)\right)\text{.}
\end{align}
So for a fixed number of sampled points $N$,
we see 
that in the unweighted case the bias of $\widehat{\Theta}^{\rm I-SIR}_{M_N}$ is independent of $M_N$.
By contrast,
whichever $N$ the bias of ${\rm E}(\widehat{\Theta}^{\rm I-SIR-w}_{M_N})$ tends to $0$ as $M_N \rightarrow \infty$.

Finally, it remains to compute $p_u(x^i)/\tilde{q}_N(x^i)$ in practice. 
In general, 
$\tilde{q}_N$ in \eqref{q-tilde-static} is not available in close form 
because it relies on the integral $h_N(x)$ in \eqref{h-static}. 
However, the $N \times M_N$ intermediate samples
which have been used in Algorithm \ref{algo-ISIR} can be recycled to approximate the 
conditional expectation $h_N(x)$. 
For a given $x$ and using the intermediate
samples $\tilde{x}^{i,j}$ of Algorithm \ref{algo-ISIR},
a crude Monte Carlo estimate of $h_N(x)$ reads
\begin{equation}
\hat{h}_N(x)= \sum_{i=1}^{M_N} \frac{\frac{p_u(x)}{q(x)}}{\frac{p_u(x)}{q(x)} + \sum_{j=1}^{N-1} \frac{p_u(\tilde{x}^{i,j})}{q(\tilde{x}^{i,j})}} \text{.}
\end{equation}
Importance weights $\frac{p_u(x)}{\tilde{q}_N(x)}$ in \eqref{theta-ISIR-2}
can be approximated by $\frac{p_u(x)}{N\hat{h}_N(x)q(x)}$. 
Note that the computation of these approximated weights do not require extra computational 
cost since  $p_u(\tilde{x}^{i,j})/q(\tilde{x}^{i,j})$ has already been computed
in Algorithm \ref{algo-ISIR} to obtain i.i.d. samples.

\subsection{Summary}
\label{summary-static}

In summary, we now have at our disposal four estimates to compute 
$\Theta$ in \eqref{Theta} from an importance distribution $q(.)$.
$\widehat{\Theta}^{\rm IS}_{N}$
and $\widehat{\Theta}^{\rm SIR}_{M_N}$  are deduced 
from the IS and Rubin's SIR mechanisms, 
respectively.
$\widehat{\Theta}^{\rm SIR}_{M_N}$ relies on unweighted 
dependent samples from $\tilde{q}_N$. Using
unweighted independent samples from $\tilde{q}_N$ produces the
estimate $\widehat{\Theta}^{\rm I-SIR}_{M_N}$ which outperforms
$\widehat{\Theta}^{\rm SIR}_{M_N}$ and possibly $\widehat{\Theta}^{\rm IS}_{N}$;
it also becomes asymptotically independent of the choice of the initial 
importance distribution $q(.)$ according to theorem \ref{prop-tcl}.
This estimate does not suffer from the support impoverishment caused by the resampling step.
On the other hand
it requires a larger computational cost which, however,
can be exploited in order to associate to the i.i.d. samples
post-resampling importance weights
based on the $\tilde{q}_N(x)$. 
We thus obtain a weighted estimate $\widehat{\Theta}^{\rm I-SIR-w}_{M_N}$ 
which can be 
seen as the estimate deduced from the IS mechanism 
based on the compound IS distribution $\tilde{q}_N(x)$.
We will compare these estimates via simulations 
and will take into account their computational cost in Section \ref{simu-static}.


\section{Independent resampling based PF}
\label{section-2}

We now adapt the results of Section \ref{section-1} 
to the Bayesian filtering problem.
In section \ref{dynamique-section-sir} 
we briefly recall the principle of classical 
SIR algorithms which are based on dependent resampling.
Our SIR algorithm with independent resampling and unweighted samples
is proposed in section \ref{dynamique-section-sir-ind}.
However, computing the post-resampling weights is more challenging 
here than in the static case because 
the pdf $\tilde{q}_N$ of the static case 
becomes a sum of $N$ terms which should be computed for each final sample.
So in section \ref{section-connection-APF}
we revisit the algorithm of section \ref{dynamique-section-sir-ind} in terms of APF.
We first observe that the independent SIR algorithm 
can be seen as the first step of an APF algorithm
since it implicitly draws samples from a mixture pdf. 
Making full use of the APF methodology 
enables us to weight our final samples.

\subsection{Classical SIR algorithms (based on dependent resampling)} 
\label{dynamique-section-sir}

We now assume that we are given some hidden Markov model \eqref{HMC}
and we briefly recall how 
$\Theta_k$ in \eqref{Theta-HMC}
can be computed recursively via PF.
PF relies on the sequential application of the normalized
IS mechanism described in Section \ref{review-SIR} for the target
distribution $p(x_{0:k}|y_{0:k})$ which is known up to a constant
according to \eqref{HMC}.
Let $q(x_{0:k})$ be an importance distribution 
($q(x_{0:k})$ can depend on $y_{0:k}$ but this dependency is not written here to avoid notational burden).
Starting from $N$ weighted trajectories 
$x_{0:k-1}^i$ sampled from $q(x_{0:k-1})$, 
we first extend each trajectory $x_{0:k-1}^i$ by a particle $\tilde{x}_k^i$ sampled from $q(x_k|x_{0:k-1}^i)$
and next update the old weights $w_{k-1}^i$ via  
\begin{equation}
\label{poids}
w_k^i \propto w_{k-1}^i \frac{f_{k}(\tilde{x}_k^i|x_{k-1}^i)g_k(y_k|\tilde{x}_k^i)}{q(\tilde{x}_k^i|x_{0:k-1}^i)}, \sum_{i=1}^N w_k^i =1 \text{.}
\end{equation}
Unfortunately, it is well-known that this direct sequential application of IS 
leads to weight degeneracy:
after a few iterations only few weights $w_k^i$ have a non null value \cite{doucet-sequentialMC}.
A traditional rescue consists in resampling, 
either systematically or according to some criterion such as 
the Effective Sample Size \cite{kong1994} \cite{liu-chen1995}
which is approximated by $1/\sum_{i=1}^N (w_k^i)^2$.
The corresponding algorithm is given in Algorithm
\ref{algo-PFSIR} and we shall assume 
that the size $N$ of the MC approximation 
remains constant thoughout the iterations.
Finally Algorithm \ref{algo-PFSIR} enables to compute 
two estimates of $\Theta_k$:
\begin{eqnarray}
\label{theta-sis-seq}
\widehat{\Theta}_{N,k}^{\rm SIS}&=& \sum_{i=1}^N w_k^i f(\tilde{X}_k^i) \text{,}\\
\label{theta-sir-seq}
\widehat{\Theta}_{N,k}^{\rm SIR}&=&\frac{1}{N} \sum_{i=1}^N f(X_k^i) \text{.}
\end{eqnarray}
As is well known, 
the pre-resampling estimator 
$\widehat{\Theta}_{N,k}^{\rm SIS}$ is preferable to 
the post-resampling one
$\widehat{\Theta}_{N,k}^{\rm SIR}$ and should be used in practice;
but $\widehat{\Theta}_{N,k}^{\rm SIR}$ is recalled here 
because it will be compared below to the independent resampling estimator 
\eqref{theta-isir-seq}.

\begin{algorithm}
\SetKwInOut{KwInput}{Input}
\KwInput{$q(x_k|x_{0:k-1})$, $y_k$, $\{w_{k-1}^i,x_{0:k-1}^i\}_{i=1}^{N}$}
\KwResult {$\{w_k^i,x_{0:k}^i\}_{i=1}^{N}$}
\For{$1 \leq i \leq N$} 
{
\textbf{S.} $\tilde{x_k}^{i} \sim q(x_k|x_{0:k-1}^i)$;  \\
\textbf{W.} $w_k^{i} \propto  w_{k-1}^i\frac{f_{k}(\tilde{x}_k^i|x_{k-1}^i)g_k(y_k|\tilde{x}_k^i)}{q(\tilde{x}_k^i|x_{0:k-1}^i)}$ \text{, } $\sum_{i=1}^N w_k^i =1$; \\
}
\eIf{Resampling}
{
\For{$1 \leq i \leq N$} {
\textbf{R.} $l^i \sim {\rm Pr}(L=l|\{x_{0:k-1}^j,\tilde{x}_k^j\}_{j=1}^N)=w_k^l $ \\
Set $x_{0:k}^i=(x_{0:k-1}^{l^i},\tilde{x}_k^{l^i})$, $w_k^i=\frac{1}{N}$
}
}
{
Set $\{x_k^i\}_{i=1}^N=\{\tilde{x}_k^i\}_{i=1}^N$
}
\caption{The classical SIR algorithm (based on dependent resampling)}
\label{algo-PFSIR}
\end{algorithm}


In practice, 
it remains to choose the conditional importance distribution 
$q(x_k|x_{0:k-1})$. A popular solution consists in choosing 
$q(x_k|x_{0:k-1})=f_{k}(x_k|x_{k-1})$,
since this pdf is part of  model \eqref{HMC} and is generally 
easy to sample from; 
another one is the so-called optimal
conditional importance distribution $q(x_k|x_{0:k-1})=p(x_k|x_{k-1},y_k)$
which takes into account the new observation $y_k$ 
and for which 
weights $w_k^i$ no longer depend on the sampled particles
$\{\tilde{x}_k^i\}_{i=1}^N$. 
The optimal conditional
importance distribution is generally not available in closed form 
but some approximation
techniques have been proposed, 
see e.g. 
\cite{doucet-sequentialMC}
\cite{Merwe_UPF} 
\cite{Saha_EMM}. The choice of the importance distribution
will be not discussed in this paper and does not impact the proposed methodology.
Finally, let us mention that convergence results are also available for the
PF presented in Algorithm \ref{algo-PFSIR}, see e.g.
\cite{crisan-doucet} \cite{chopin} \cite{delmoral-livre} \cite{Cappeetal}.
Some of them are based on the recursive application of the CLTs 
recalled in Section \ref{section-1}.

\subsection{An alternative SIR algorithm (based on independent resampling)}
\label{dynamique-section-sir-ind}

Let us first adapt Proposition \ref{prop-q-tilde-static} to the sequential context.
So we address the conditional distribution given $\{x_{0:k-1}^i\}_{i=1}^N$
of the resampled particles $X_k^i$ 
and we have the following result (the proof is omitted).

\begin{proposition}
Let us consider the samples $\{X_k^i\}_{i=1}^{N}$ produced by 
the SIR mechanism of Algorithm \ref{algo-PFSIR}.
Let
\begin{eqnarray}
p_{i,k}(x)&=&w_{k-1}^if_{k}(x|x_{k-1}^i)g_k(y_k|x) \text{,} \\
q_{i,k}(x)&=&q(x|x_{0:k-1}^i) \text{.}
\end{eqnarray}
Then given the initial trajectories $\{x_{0:k-1}^i\}_{i=1}^{N}$,
the new samples $\{X_k^i\}_{i=1}^{N}$ are 
identically distributed according to a pdf 
$\tilde{q}_{N,k}$ which reads
\begin{eqnarray}
\label{q-tilde-dynamic}
\tilde{q}_{N,k}(x)= \sum_{i=1}^N h_{i,k}(x) q_{i,k}(x) \text{,} \end{eqnarray}
where $h_{i,k}(x)$ coincides with the conditional expectation (given $(X_k^i=x)$) of
the $i$-th importance weight at time $k$, 
\begin{equation}
\label{h-dynamic}
h_{i,k}(x)= \int \int \frac{\frac{p{i,k}(x)}{q_{i,k}(x)}}{\frac{p_{i,k}(x)}{q_{i,k}(x)} + \sum_{l\neq i} \frac{p_{l,k}(x^l)}{q_{l,k}(x^l)}} \prod_{l\neq i} q_{l,k}(x^l) {\rm d}x^l \text{.}
\end{equation}
\end{proposition}
Note that in this proposition we focus on the distribution 
of $X_k^i$ given $\{x_{0:k-1}^i\}_{i=1}^{N}$.
Given $\{x_{0:k-1}^i, \tilde{x}_k^i\}_{i=1}^{N}$,
the new samples $\{X_k^i\}_{i=1}^{N}$ are independent; 
when we remove the dependency in $\{\tilde{x}_k^i\}_{i=1}^{N}$,
$\{X_k^i\}_{i=1}^{N}$ become  identically distributed according to $\tilde{q}_{N,k}$ 
but are dependent 
(a same particle can be resampled several times).

Since $\tilde{q}_{N,k}$ is a pdf, a procedure which would produce 
samples conditionally i.i.d. from $\tilde{q}_{N,k}$
would enable us to keep the advantage of the resampling step, 
i.e. to recreate diversity for the next time iteration 
while avoiding local impoverishment of the support. 
Except in a particular case which will be described later, 
sampling directly from $\tilde{q}_{N,k}(x)$ is difficult 
for an arbitrary conditional importance distribution $q(x_k|x_{0:k-1})$.
We thus propose a procedure similar to Algorithm 
\ref{algo-ISIR} but adapted to the dynamical context. 
The SIR algorithm with independent resampling is given by Algorithm \ref{algo-PFSIR-ind}.
Note that a difference with Algorithm \ref{algo-PFSIR} 
is that the distribution of the discrete index $L^i$ now depends on $i$.
 
\begin{algorithm}
\SetKwInOut{KwInput}{Input}
\KwInput{$q(x_k|x_{0:k-1})$, $y_k$, $\{w_{k-1}^i,x_{0:k-1}^i\}_{i=1}^{N}$}
\KwResult {$\{w_k^i,x_{0:k}^i\}_{i=1}^{N}$}
\For{$1 \leq i \leq N$} 
{
\For{$1 \leq j \leq N$} 
{
\textbf{S.} $\tilde{x_k}^{i,j} \sim q(x_k|x_{0:k-1}^i)$;  \\
\textbf{W.} $w_k^{i,j} \propto  \frac{f_{k}(\tilde{x}_k^{i,j}|x_{k-1}^i)g_k(y_k|\tilde{x}_k^{i,j})}{q(\tilde{x}_k^{i,j}|x_{0:k-1}^i)}$, $\sum_{j=1}^N w_k^{i,j}=1$;
}
\textbf{R.} $l^i \sim {\rm Pr}(L=l | \{x_{0:n-1}^j,\tilde{x}_n^{i,j}\}_{j=1}^N)=w_k^{i,l}$ \\
Set $x_{0:k}^i=(x_{0:k-1}^{l^i},\tilde{x}_k^{i,l^i})$, $w_k^i=\frac{1}{N}$
}
\caption{A SIR algorithm based on independent resampling}
\label{algo-PFSIR-ind}
\end{algorithm}

We now propose a new estimate 
$\widehat{\Theta}_{N,k}^{\rm I-SIR}$ of $\Theta$
which is based on the set $\{X_k^i\}_{i=1}^N$ produced by Algorithm \ref{algo-PFSIR-ind}:
\begin{equation}
\label{theta-isir-seq}
\widehat{\Theta}_{N,k}^{\rm I-SIR}=\frac{1}{N} f(X_k^i) \text{.}
\end{equation}
Comparing 
\eqref{theta-isir-seq}
with 
\eqref{theta-sir-seq},
remember that the samples $\{X_k^i\}_{i=1}^N$ share the same pdf $\tilde{q}_{N,k}$,
but that in \eqref{theta-isir-seq} they are now independent given $\{x_{0:k-1}^i\}_{i=1}^N$.
Starting from a dataset $\{x_{0:k-1}^i\}_{i=1}^N$,
it is ensured that $\widehat{\Theta}_{N,k}^{\rm I-SIR}$ outperforms $\widehat{\Theta}_k^{\rm SIR}$ since
\begin{align}
{\rm E}(\widehat{\Theta}_{N,k}^{\rm I-SIR}|\{x_{0:k-1}^i\}_{i=1}^N) &= {\rm E}(\widehat{\Theta}_n^{\rm SIR}|\{x_{0:k-1}^i\}_{i=1}^N) \text{,} \\
{\rm var}(\widehat{\Theta}_{N,k}^{\rm SIR}|\{x_{0:k-1}^i\}_{i=1}^N) &= {\rm var}(\widehat{\Theta}_n^{\rm I-SIR}|\{x_{0:k-1}^i\}_{i=1}^N) +
\frac{N-1}{N}{\rm var}(\widehat{\Theta}_{N,k}^{\rm SIS}|\{x_{0:k-1}^i\}_{i=1}^N) \text{.}
\end{align}
Of course, computing $\widehat{\Theta}_{N,k}^{\rm I-SIR}$ 
via the samples produced by Algorithm \ref{algo-PFSIR-ind}
requires an extra computational cost. 
This point will be discussed in detail in our Simulations section, 
but for the moment let us make two comments:
first, this algorithm can be seen as an alternative resampling scheme 
which ensures the diversity of the resampled support without changing 
the conditional distribution of the final samples; 
if resampling needs to be performed rarely, 
then the independent resampling procedure may be used only when necessary. 
On the other hand, 
we will see that
$\widehat{\Theta}_{N,k}^{\rm I-SIR}$
can also provide an interesting alternative 
to $\widehat{\Theta}_{N,k}^{\rm SIS}$ but requires
an extra computational cost; 
so if we want to perform the independent resampling procedure at each time step
we will decrease the number $N$ of particles associated with
$\widehat{\Theta}_{N,k}^{\rm I-SIR}$ in order 
to reach the same computational cost associated with $\widehat{\Theta}_{N,k}^{\rm SIS}$.

\begin{remark}
Note that the idea of using extra MC samples 
has already been proposed in the context of Island PFs  
\cite{Verge2015}.
The idea behind this class of techniques 
is to exploit parallel architectures, 
and the rationale is as follows. 
Instead of considering a unique set of $N$ particles, 
the method consists in dividing the population of
$N$ samples into $N_1$ sets of $N_2$ samples
such as $N_1 N_2=N$. 
It is well known that
such a configuration does not improve the
classical PF with $N$ samples, 
but it has the advantage
to split the associated computational cost
when parallel architectures are available. 
In other words, the objective of the PFs is
not to struggle against the support impoverishment.
\end{remark}


\subsection{Interpretation of the independent sampling scheme in terms of APF}
\label{section-connection-APF}

At this point, 
we have seen that it was possible to obtain an estimate of $\Theta_{k}$ 
based on i.i.d. samples from the conditional pdf $\tilde{q}_{N,k}$. 
As in the static case,
we now wonder whether the final weights $1/N$ used to compute
$\widehat{\Theta}_{N,k}^{\rm I-SIR}$ (see eq. 
\eqref{theta-isir-seq})
are optimal when $N$ is finite. 
To this end we would like to make use of 
the expression of $\tilde{q}_{N,k}$ 
to propose an alternative weighting mechanism.
At first glance, the computation of a weight which would
rely on 
\eqref{q-tilde-dynamic}-\eqref{h-dynamic}
seems compromised because
$\tilde{q}_{N,k}$ 
involves a sum of $N$ terms which should be computed
for each $N$ final sample $x_k^i$.
As we will see, 
the interpretation of the independent SIR algorithm
as a particular first step of an APF algorithm
will help circumvent this limitation.
Let us first begin with a brief presentation of APF filters.

\subsubsection{A brief presentation of APF}
\label{presentation-APF}

In model \eqref{HMC}, 
the filtering density at time $k$ can be
written in terms of that at time $k-1$,
\begin{equation}
\label{filtrage}
p(x_k|y_{0:k}) \propto g_k(y_k|x_k) \int f_{k}(x_k|x_{k-1}) p(x_{k-1}|y_{0:k-1}) {\rm d}x_{k-1} \text{.}
\end{equation}
Plugging an MC approximation $\{w_{k-1}^i,x_{k-1}^i\}_{i=1}^N$ 
of $p(x_{k-1}|y_{0:k-1})$ into \eqref{filtrage} yields
\begin{eqnarray}
\nonumber
\widehat{p}(x_k|y_{0:k}) &\propto& g_k(y_k|x_k) 
\sum_{i=1}^N w_{k-1}^i f_{k}(x_k|x_{k-1}^i) \text{,} \\
\label{filtrage-3}
 &\propto& \sum _{i=1}^N w_{k-1}^i p(y_k|x_{k-1}^i) p(x_k|x_{k-1}^i,y_{k}) \text{,} 
\end{eqnarray}
where
$p(y_k|x_{k-1})= \int f_{k}(x_k|x_{k-1})g_k(y_k|x_k) {\rm d}x_{k-1}$
and 
$p(x_k|x_{k-1},y_k)  \propto  f_{k}(x_k|x_{k-1})g_k(y_k|x_k)$.
Sampling from $\widehat{p}(x_k|y_{0:k})$ in \eqref{filtrage-3}
leads to a particular SMC algorithm
refered to as the FA-APF \cite{auxiliary}.
However
sampling directly from $\widehat{p}(x_k|y_{0:k})$ is not necessarily possible
because $p(y_k|x_{k-1}^i)$ or $p(x_k|x_{k-1},y_k)$ are often unavailable.
To that end it has been proposed \cite{auxiliary} 
to obtain samples 
from an instrumental mixture pdf
\begin{equation}
\label{mixture-IS}
\overline{q}(x_k)=\sum_{i=1}^N \mu(x_{0:k-1}^i)\tau(x_k|x_{0:k-1}^i)
\end{equation}  
and to use IS in augmented dimension;
finally APF aims at targeting the mixture pdf $\widehat{p}(x_k|y_{0:k})$ in \eqref{filtrage-3} 
which, itself, 
targets the filtering distribution $p(x_k|y_{0:k})$.
The resulting algorithm is displayed below.

\begin{algorithm}
\SetKwInOut{KwInput}{Input}
\KwInput{$\mu(x_{0:k-1})$, $\tau(x_k|x_{0:k-1})$, $y_k$, $\{w_{k-1}^i,x_{0:k-1}^i\}_{i=1}^{N}$}
\KwResult {$\{w_k^i,x_{0:k}^i\}_{i=1}^{N}$}
\For{$1 \leq i \leq N$} 
{
\textbf{R.} $l^i \sim {\rm Pr}(L=l| \{x_{0:k-1}^i\}_{i=1}^{N}) = \mu(x_{0:k-1}^l)$\\
\textbf{S.} $x_k^{i} \sim \tau(x_k|x_{0:k-1}^{l^i})$;  \\
\textbf{W.} $w_k^{i} \propto  \frac{w_{k-1}^{l^i} f_{k}(x_k^i|x_{k-1}^{l^i})g_k(y_k|x_k^i)}{\mu(x_{0:k-1}^{l^i})\tau(x_k^i|x_{0:k-1}^{l^i})}$, $\sum_{i=1}^N w_k^i =1$; \\
Set $x_{0:k}^i=(x_{0:k-1}^{l^i},x_k^i)$
}
\caption{The APF algorithm}
\label{algo-APF}
\end{algorithm}

Let us comment the choice of the instrumental distribution $\overline{q}(x_k)$ in
\eqref{mixture-IS}.
Compared to the SIS algorithm of paragraph \ref{dynamique-section-sir}
we see that there is an additional degree of freedom, $\mu(x_{0:k-1})$, 
which is called the first stage weight;
$\tau(x_k|x_{0:k-1}^i)$ refers to a given conditional importance distribution.
Generally, the objective of the first stage weights is 
to avoid the computational waste induced by the resampling step of the SIR algorithm 
by pre-selecting trajectories at time $k-1$ 
which are in accordance with the new observation $y_k$.
Designing this pdf $\overline{q}(x_k)$ is critical 
and classical approximations of the predictive likelihood such as 
the likelihood taken at the mode of the transition pdf
(i.e. $\mu(x_{0:k-1}^i) \propto w_{k-1}^i g_k(y_k|\phi(x_{k-1}^i))$ where
$\phi(x_{k-1}^i)$ is the mode of $f_{k|k-1}(x_k|x_{k-1}^i)$) can actually damage the 
performance of the estimate.
This is why it is often suggested in practice to build a first-stage weight as close as possible to $w_{k-1}p(y_k|x_{k-1})$,
although this problem is generally difficult \cite{doucet-APF} \cite{whiteley-johansen}
due to the computation of the predictive likelihood $p(y_k|x_{k-1})$.
It remains to choose the importance distribution $\tau(x_k|x_{0:k-1})$;
as in the SIR algorithm, 
one generally tries to approximate the optimal importance distribution $p(x_k|x_{k-1},y_k)$. 
Finally note that similarly to classical IS, 
the FA-APF setting is not necessarily optimal from an asymptotic point of view 
even if it performs very well in practice \cite{douc-APF}.

\subsubsection{Independent resampling as the first step of a canonical APF algorithm}
\label{paragraph-interpretation}

Let us now turn to the interpretation of our independent resampling procedure
in terms of APF. 
Let us observe that $\tilde{q}_{N,k}$ in \eqref{q-tilde-dynamic} 
can be rewritten as
\begin{equation}
\label{mixture-q-tilde}
\tilde{q}_{N,k}(x)= \sum_{i=1}^N \int h_{i,k}(x)q_{i,k}(x){\rm d}x \times \frac{h_{i,k}(x)q_{i,k}(x)}{\int h_{i,k}(x)q_{i,k}(x){\rm d}x} \text{}
\end{equation}
and so can be seen as one particular mixture pdf
$\overline{q}(x_k)$ in \eqref{mixture-IS},
in which the weights 
$\mu^{\rm ind}(x_{0:k-1}^i)$ are given by $\int h_{i,k}(x)q_{i,k}(x){\rm d}x$ 
and the components $\tau^{\rm ind}(x_k|x_{0:k-1}^i)$ by $\frac{h_{i,k}(x)q_{i,k}(x)}{\int h_{i,k}(x)q_{i,k}(x){\rm d}x}$.
We now verify that the couple of samples $(l^i,x_k^i)$ produced by the
independent resampling algorithm (Algorithm \ref{algo-PFSIR-ind})
can indeed be seen as an augmented sample according to 
$\tilde{q}_{N,k}(x)$ in 
\eqref{mixture-q-tilde}:
\begin{itemize}
\item given $\{x_{0:k-1}^j\}_{j=1}^N$ and $\{\tilde{x}_k^{i,j}\}_{j=1}^N$, 
${\rm Pr}(L^i=l)=w_k^{i,l}$. Since $\tilde{x}_k^{i,j}\sim q_i(x)$,
the distribution of $l^i$ given $\{x_{0:k-1}^j\}_{j=1}^N$ 
becomes ${\rm Pr}(L^i=l)={\rm E}(w_k^l|\{x_{0:k-1}^j\}_{j=1}^N) =\int h_{l,k}(x)q_{l,k}(x){\rm d}x$;
\item given $\{x_{0:k-1}^j\}_{j=1}^N$, $\{\tilde{x}_k^{i,j}\}_{j=1}^N$ and
$l^i$, $x_k^i=\tilde{x}_k^{i,l^i}$.  Removing the dependency in $\{\tilde{x}_k^{i,j}\}_{j=1}^N$,
the distribution of $x_k^i$
given $\{x_{0:k-1}^j\}_{j=1}^N$ and $l^i$ becomes 
$\frac{h_{l^i,k}(x)q_{l^i,k}(x)}{\int h_{l^i,k}(x)q_{l^i,k}(x){\rm d}x}$.
\end{itemize}
In summary,
our independent resampling procedure 
is nothing but the first step of one particular APF algorithm,
because the pdf
$\tilde{q}_{N,k}(x)$
from which we draw i.i.d. samples (given $\{w_{k-1}^i,x_{0:k-1}^i\}_{i=1}^N$)
coincides with the mixture pdf \eqref{mixture-q-tilde},
which itself constitutes a class of instrumental distributions 
$\overline{q}(x_k)$ in \eqref{mixture-IS} parametrized by $q(x_k|x_{0:k})$.

In order to appreciate the relevance of that particular solution 
let us comment on the choice 
of the first-stage weights $\mu^{\rm ind}(x_{0:k-1}^i)$ 
and distributions $\tau^{\rm ind}(x_k|x_{0:k-1}^i)$:
\begin{itemize}
\item at time $k-1$,
trajectories $\{x_{0:k-1}^i\}_{i=1}^N$ are first
resampled according to a first stage weight which coincides 
with the expectation of the importance weights $w_k^i$ of the SIR algorithm
defined in \eqref{poids}. 
In other words, these
trajectories are preselected 
in such a way that the new importance weight $w_k^i$
which would be affected in the weighting step of the SIR algorithm will tend to be large;
\item once a trajectory $x_{0:k-1}^i$ has been selected, 
it is not ensured that its associated weight $w_k^i$ will indeed be large.
By sampling according to a pdf proportional to $h_{i,k}(x)q_{i,k}(x)$, 
the objective is to produce a sample in the region where $h_{i,k}(x)$ 
(the conditional expectation of the importance weight $w_{k}^i$, given that $(X_k^i=x$)) and the distribution $q_{i,k}(x)$ are large.
\end{itemize}
Consequently, the mixture pdf $\tilde{q}_{N,k}(x)$ 
appears as a natural instrumental candidate for the APF 
when the objective is to pre-select the trajectories 
and to extend them in accordance with the given
conditional importance distributions $q_{i,k}(x)=q(x|x_{0:k-1}^i)$
used in the SIR algorithm.
If the SIR algorithm IS densities $q_{i,k}(x)$ 
coincide with the optimal importance distribution $p(x|x_{k-1}^i,y_k)$,
then one can see easily that our canonical APF instrumental pdf 
\eqref{mixture-q-tilde} reduces to the target mixture \eqref{filtrage-3}
(since $h_{k,i}$ in \eqref{h-dynamic} is reduced to a term proportional 
to $w_{k-1}^i p(y_k|x_{k-1}^i)$)
and the independent SIR procedure to the FA-APF algorithm.
In that case one can sample from $\tilde{q}_{N,k}$ very efficiently
(since \eqref{mixture-q-tilde} is a known mixture)
and the resulting estimate outperforms the SIR estimate 
$\widehat{\Theta}_{N,k}^{\rm SIR}$ 
with optimal conditional importance distribution \cite{Cappeetal} \cite{petetin-sco}.
In the case where the FA-APF algorithm is not available, 
it remains possible to sample from the mixture pdf $\tilde{q}_{N,k}(x)$ in \eqref{mixture-q-tilde}
as soon as we can sample from the root pdf $q_{i,k}(x)$,
even when 
$\mu^{\rm ind}(x_{0:k-1}^i)$ cannot be computed, 
or one cannot sample from $\tau^{\rm ind}(x_k|x_{0:k-1}^i)$.


\subsubsection{Reweighting the independent samples?}
\label{reweighting-dynamique}

We can finally use this APF interpretation in order to
reweight our conditional independent samples
$\{x_{k}^i\}_{i=1}^N$.
Since $\tilde{q}_{N,k}$ can be seen as a mixture \eqref{mixture-IS}
with parameters 
$\mu^{\rm ind}(x_{0:k-1}^i)$ and $\tau^{\rm ind}(x_k|x_{0:k-1}^i)$,
$\mu^{\rm ind}(x_{0:k-1}^i) \times \tau^{\rm ind} (x|x_{0:k-1}^i)$ 
reduces to
$h_{i,k}(x)q_{i,k}(x)$.
Finally when we target 
mixture \eqref{filtrage-3},
the second-stage weights associated with
the independent samples $x_k^i$ produced by Algorithm
\ref{algo-PFSIR-ind} read
\begin{eqnarray}
\label{poids-ind}
w_k^i \propto \frac{w_{k-1}^{l^i} f_{k|k-1}(x_k^i|x_{k-1}^{l^i})g_k(y_k|x_k^i)}{h_{l^i,k}(x_k^i) q_{l^i,k}(x_k^i)}, \sum_{i=1}^N w_k^i =1.
\end{eqnarray} 
We thus obtain a new estimate of $\Theta_k$,
\begin{eqnarray}
\label{Theta-ind-2}
\widehat{\Theta}^{\rm I-SIR-w}_{N,k}= \sum_{i=1}^N w_k^i f(x_k^i)
\end{eqnarray}
where $w_k^i$ are defined in \eqref{poids-ind}. 
The practical computation of these final weights
relies on that of $h_{i,k}(x)$ in \eqref{h-dynamic},
which can be approximated by recycling the extra samples 
$\tilde{x}^{i,j}$ generated 
in Algorithm \ref{algo-PFSIR-ind}, 
\begin{equation}
\label{poids-approche}
\hat{h}_l(x)=\sum_{i=1}^N \frac{\frac{p_{l,k}(x)}{q_{l,k}(x)}}{\frac{p_{l,k}(x)}{q_{l,k}(x)}+\sum_{j\neq l} \frac{p_{j,k}(\tilde{x}^{i,j})}{q_{j,k}(\tilde{x}^{i,j})}} \text{.}
\end{equation}

\subsection{Summary} 
Let us summarize the discussions of section \ref{section-2}.
When the objective is to compute $\Theta_k$ in \eqref{Theta-HMC}
we have several options:
\begin{enumerate}
\item 
using the classical SIR algorithm (see Algorithm \ref{algo-PFSIR})
in which we compute $\widehat{\Theta}^{\rm SIS}_{N,k}$ 
defined in \eqref{theta-sis-seq}. 
The resampling step which follows the computation of this estimate
produces a conditionally dependent 
unweighted set of particles
sampled from $\tilde{q}_{N,k}$;
\item 
an alternative to avoid the local impoverishment 
induced by the traditional resampling step 
is to perform Algorithm \ref{algo-PFSIR-ind} and to compute estimate 
$\widehat{\Theta}^{\rm I-SIR}_{N,k}$. 
This estimate is still based on an unweighted set of particles marginally sampled from $\tilde{q}_{N,k}$
but these samples have become conditionally independent;
\item finally, the samples produced by Algorithm \ref{algo-PFSIR-ind} 
can also be seen as the result of a sampling procedure 
according to a partial APF instrumental mixture pdf \eqref{mixture-q-tilde}.
Using further the APF methodology
with mixture $\tilde{q}_{N,k}$ 
it is possible to target mixture \eqref{filtrage-3} 
which itself is an approximation of $p(x_k|y_{0:k})$.
This leads to estimate $\widehat{\Theta}^{\rm I-SIR-w}_{N,k}$ in \eqref{Theta-ind-2},
in which the weights \eqref{poids-ind}
are estimated by recycling the extra samples produced by Algorithm \ref{algo-PFSIR-ind}. 
\end{enumerate}
These three estimates are now going to be compared (in terms of performances and computational cost) in the next section.

\section{Simulations}   
\label{section-3}

We now validate our discussions through computer-generated experiments.
In section \ref{simu-static} we first illustrate the results of Section \ref{section-1} 
and we compare the classical resampling mechanism 
to the independent one with both unweighted and weighted samples. 
We also discuss the computational cost associated
with our independent resampling mechanism.

In section \ref{simu-APF}
we next perform simulations in the ARCH model.
On the one hand, the FA-APF algorithm can be computed in this model \cite{auxiliary}. 
On the other hand, 
remember that our weighted estimate \eqref{Theta-ind-2} 
can be interpreted as the estimate deduced from a particular
APF which uses the instrumental mixture pdf $\tilde{q}_{N,k}$ in \eqref{mixture-q-tilde},
from which it is always possible to sample from
(with an extra computational cost). 
Thus the estimate deduced from the FA-APF algorithm
is used as a benchmark and enables us to analyze the relevance
of the instrumental pdf $\tilde{q}_{N,k}$ in the APF algorithm.


Next in section \ref{simu-polar}
we compute our independent estimates for a target tracking
problem with range-bearing measurements. 
Our estimates are compared to those obtained from the classical SIR algorithm,
for a given computational budget
measured via the number of sampling operations;
this means that we compare 
$\widehat{\Theta}^{\rm I-SIR}_{M,k}$ and
$\widehat{\Theta}^{\rm I-SIR-w}_{M,k}$ 
($M$ is the number
of particles after the independent resampling step) to
$\widehat{\Theta}^{\rm SIS}_{N,k}$ in which 
$N=\frac{M^2+M}{2}$. 
Thus all estimates are based on $M^2+M$ sampling operations 
(we do not distinguish if we sample according to a continuous or a discrete distribution). 
The relative performances
of the estimates are analyzed in function of the parameters of the 
state-space model.

Finally in section \ref{simu-highdim}
we compute our estimates in models where 
the dimension $m$ of the hidden state is large and 
we analyze their performances w.r.t. classical
PF estimates in function of the dimension $m$ 
and with a fixed number of sampling operations. 
%
Finally throughout this section 
our simulations are averaged over $P=1000$ MC runs,
we set $f(x)=x$ in \eqref{Theta-HMC}
and we use an averaged Root Mean Square Error (RMSE) criterion,
defined as 
\begin{equation}
\label{RMSE}
{\rm RMSE}(\widehat{\Theta})=\frac{1}{T}\sum_{k=1}^T \left(\frac{1}{P}\sum_{p=1}^P ||\widehat{\Theta}_{k,p}-x_{k,p}||^2 \right)^{1/2}
\end{equation}
where $x_{k,p}$ is the true state at time $k$ for the $p$-th realization,
$\widehat{\Theta}_{k,p}$ is an estimate of $x_{k,p}$ and $T$ is the time length 
of the scenario.


\subsection{Comparison of static sampling procedures}
\label{simu-static}
Let us first consider the (static) Bayesian 
estimation problem in which we look for computing
\begin{equation}
\Theta = {\rm E}(X|y) = \int xp(x|y) {\rm d}x 
\end{equation} 
via the techniques described in Section \ref{section-1}.
We assume that $p(x|y)$ is known up to a constant, 
$p(x|y)\propto p(x)p(y|x)$ where $p(x)=\mathcal{N}(x;0;\sigma^2_x)$ 
and $p(y|x) = \mathcal{N}(y;x,\sigma_y^2)$ with 
$\sigma_x^2=10$ and  $\sigma_y^2=3$.
We chose the IS distribution $q(x)=p(x)$.
For a given number of final samples $N$, we compute six estimates:
the estimate $\widehat{\Theta}_{N}^{\rm SIS}$
deduced from the IS mechanism with importance distribution $q(.)$;
the estimate $\widehat{\Theta}_{N}^{\rm SIR}$
deduced from the
SIR mechanism with $N$ intermediate samples and $M_N=N$ final samples;
our estimate $\widehat{\Theta}_{N}^{\rm I-SIR}$ based on $N$ unweighted 
independent samples drawn from $\tilde{q}_N$ (see \eqref{theta-isir}); 
our estimate $\widehat{\Theta}_{N}^{\rm I-SIR-w}$ based on $N$ weighted 
independent samples from $\tilde{q}_N$ (see \eqref{theta-ISIR-2}).
Remember that the computation of the independent resampling mechanism is based
on the sampling of $N^2$ intermediate particles and 
$N$ resampling steps and thus requires an extra computational
cost w.r.t. the dependent one.
Consequently, we also compute $\widehat{\Theta}_{N}^{\rm SIR-2}$
based on the classical SIR procedure with $N^2$ intermediate samples
and $N$ (dependent) resampling steps; in other words this estimate
relies on $N$ dependent samples obtained from $\tilde{q}_{N^2}$.
Finally, we would like to
observe the effects of weighting the final samples in the dependent 
resampling case; so we compute  $\widehat{\Theta}_{N}^{\rm SIR-w}$ 
which relies on extra samples to approximate the weight proportional 
to $p(x,y)/\tilde{q}_{N}(x)$.

\begin{figure}[htbp!]
\centering
\includegraphics[scale=0.4]{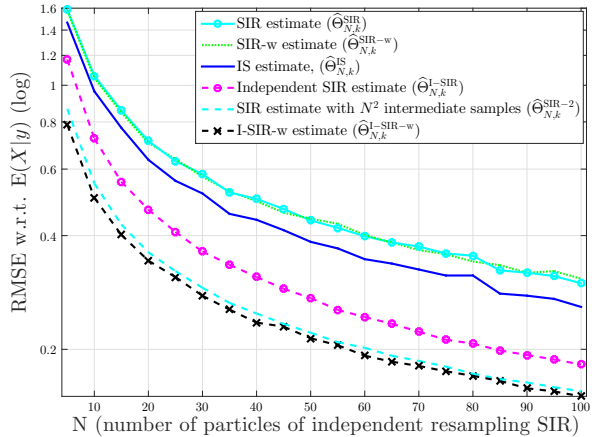}
\caption{Static linear and Gaussian model - $\sigma_x^2=10$, $\sigma_y^2=3$ -
Bayesian estimates of ${\rm E}(X|y)$ based on the independent resampling mechanism  
outperform the estimates based on the traditional IS and SIR mechanisms altough
they require an extra computational cost.
When the computational cost is fixed, the estimate based on weighted i.i.d. samples from $\tilde{q}_N$ 
outperforms the estimates based on identically distributed samples from $\tilde{q}_{N^2}$.}
\label{fig:LinearGaussianStatic}
\end{figure}

In Fig. \ref{fig:LinearGaussianStatic} we display 
the distance of each estimate w.r.t. the true expectation
${\rm E}(X|Y=y)$ in function of the number of samples $N$. 
As expected, the estimate $\widehat{\Theta}_{N}^{\rm I-SIR}$ 
based on $N$ independent samples drawn from $\tilde{q}_N$ 
outperforms the estimate $\widehat{\Theta}_{N}^{\rm SIR}$
which is computed from $N$ dependent samples drawn from $\tilde{q}_N$.
However, an interesting result is that $\widehat{\Theta}_{N}^{\rm I-SIR}$
also outperforms $\widehat{\Theta}_{N}^{\rm SIS}$. It means
that the distribution $\tilde{q}_N$ produced by the 
SIR mechanism is more adapted than the prior $q(x)=p(x)$,
which is not surprising since $\tilde{q}_N$ uses implicitly
the observation $y$ through the resampling mechanism of intermediate
samples.
Of course, the computation
of $\widehat{\Theta}_{N}^{\rm I-SIR}$ requires an extra computational
cost but it is interesting to note that the size of the final support
is the same in the three cases.
We finally compare the estimates based on the same computational cost.
When $N$ increases, these estimates have the same asymptotical 
behavior. It can be seen that the estimate 
$\widehat{\Theta}_{N}^{\rm SIR-2}$ based on $N$ samples
drawn from $\tilde{q}_{N^2}$ outperforms $\widehat{\Theta}_{N}^{\rm I-SIR}$.
However, when our i.i.d. samples are weighted by a term proportional to 
$p(x,y)/\tilde{q}_{N}(x)$ in an IS perspective,
our estimate $\widehat{\Theta}_{N}^{\rm I-SIR-w}$
has the best performance whatever $N$.
We finally note that contrary to the independent procedure,
weighting the samples when they are dependent does not improve the performance 
when compared to the estimate based on dependent and unweighted samples; 
indeed, $\widehat{\Theta}_{N}^{\rm SIR-w}$
is not any better than $\widehat{\Theta}_{N}^{\rm SIR}$.
The performances of these algorithms are also presented in terms of RMSE (w.r.t. to the true value of
$X$) in Table \ref{tbl:staticrmse}.

\begin{table}[htbp]
\centering
\begin{tabular}{|l|l|l|l|l|l|l|}
\hline
$N$                                       & $\widehat{\Theta}_{N}^{\rm SIR}$ & $\widehat{\Theta}_{N}^{\rm SIR-w}$ & $\widehat{\Theta}_{N}^{\rm SIS}$ & $\widehat{\Theta}_{N}^{\rm I-SIR}$ & $\widehat{\Theta}_{N}^{\rm SIR-2}$ & $\widehat{\Theta}_{N}^{\rm I-SIR-w}$ \\ \hline
$20$  & $1.6844$ & $1.6819$ & $1.6542$ & $1.5951$ & $1.5618$ & $\mathbf{1.5610}$ \\ \hline
$40$  & $1.5925$ & $1.5981$ & $1.5763$ & $1.5606$ & $1.5446$ & $\mathbf{1.5410}$ \\ \hline
$60$  & $1.5752$ & $1.5777$ & $1.5637$ & $1.5442$ & $1.5395$ & $\mathbf{1.5335}$ \\ \hline
$80$  & $1.5623$ & $1.5639$ & $1.5530$ & $1.5345$ & $1.5309$ & $\mathbf{1.5293}$ \\ \hline
$100$ & $1.5519$ & $1.5504$ & $1.5410$ & $1.5320$ & $\mathbf{1.5290}$ & $\mathbf{1.5290}$ \\ \hline
\end{tabular}
\caption{Static linear and Gaussian model - RMSE values of each estimate.}
\label{tbl:staticrmse}
\end{table}

\subsection{Comparison with APF algorithms}
\label{simu-APF}
We now focus on the interpretation of our independent resampling
algorithm in terms of APF.
We study the ARCH model which is a particular hidden Markov model \eqref{HMC}
in which $f_{k}(x_k|x_{k-1})=\mathcal{N}(x_k;0;\beta_{0} + \beta_{1}x^2_{k-1})$
and $g_k(y_k|x_k)=\mathcal{N}(y_k;x_k;R)$. 
We set $R=1$, $\beta_0 = 3$ and $\beta_1 = 0.75$. 
In this model one can compute
$p(y_k|x_{k-1})=\mathcal{N}(y_k;0;R + \beta_{0} + \beta_{1}x^2_{k-1})$ and 
$p(x_k|x_{k-1},y_k)=\mathcal{N}(x_k;\frac{\beta_{0} + \beta_{1}x^2_{k-1}}{R + \beta_{0} + \beta_{1}x^2_{k-1}} y_k;
\frac{R(\beta_{0} + \beta_{1}x^2_{k-1})}{R + \beta_{0} + \beta_{1}x^2_{k-1}})$;
consequently, it is possible to obtain i.i.d. samples from 
the target mixture \eqref{filtrage-3} and thus to compute
the estimate $\widehat{\Theta}^{\rm FA}_{N,k}$ based on the FA-APF
algorithm. Remember that the FA-APF can also be 
seen as a particular case of our independent resampling
Algorithm \ref{algo-PFSIR-ind} in which the importance distribution $q(x_k|x_{0:k-1})$
coincides with $p(x_k|x_{k-1},y_k)$ 
(see section \ref{paragraph-interpretation}).
However this setting can be implemented in specific models only,
while Algorithm \ref{algo-PFSIR-ind} can be used 
with any importance distribution $q(x_k|x_{0:k-1})$,
while keeping the same interpretation as the FA-APF 
(see our discussion in section \ref{paragraph-interpretation}). 
So we also compute our estimates $\widehat{\Theta}^{\rm I-SIR}_{N,k}$
and $\widehat{\Theta}^{\rm I-SIR-w}_{N,k}$ which can be seen 
as an estimate deduced from the APF in
which the importance mixture \eqref{mixture-IS} 
coincides with $\tilde{q}_{N,k}$.
We finally compute the estimate $\widehat{\Theta}^{\rm APF}_{N,k}$
which is deduced from the APF with $\mu(x_{0:k-1})\propto w_{k-1} p(y_k|x_{k-1})$ and
$\tau(x_k|x_{0:k-1})=f_{k}(x_k|x_{k-1})$; 
with this configuration, the particles are pre-selected
with the so-called optimal first
stage weight and sampled from the transition pdf.

\begin{figure}[htbp]
\centering
\begin{subfigure}[RMSE]
{
\centering
\includegraphics[scale=0.4]{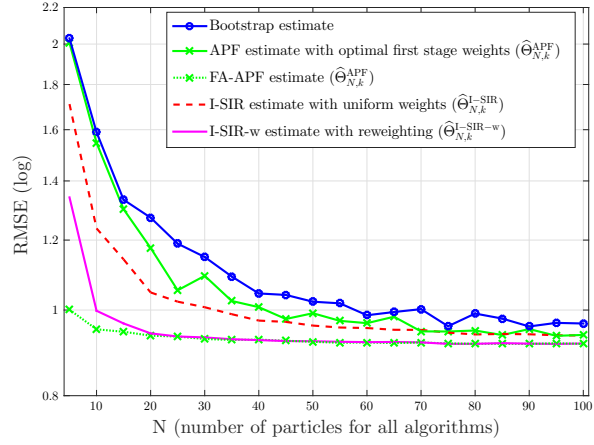}
\label{fig:APFComparisonARCH}
}
\end{subfigure}

\begin{subfigure}[Averaged normalized effective sample size]
{
\centering
\includegraphics[scale=0.4]{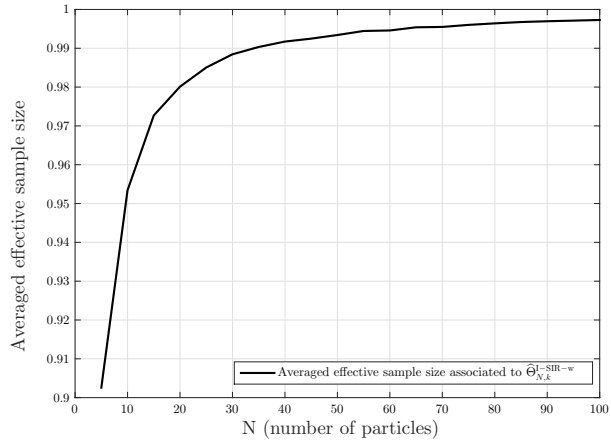}
\label{fig:CvgPQtildeARCH}
}
\end{subfigure}
\caption{ARCH model - $R=1$, $\beta_0 = 3$ and
$\beta_1 = 0.75$ - (a) The estimate based on the independent resampling mechanism with a final reweighting
has the same performances as the estimate deduced from the FA-APF. The final reweighting mechanism 
is beneficial when compared to the use of uniform weighs - (b) When $N$ is large, final weights associated
to the estimate $\widehat{\Theta}^{\rm I-SIR-w}_{N,k}$ tend to be uniform.}
\end{figure}

The RMSE of each estimate 
is displayed in Fig. \ref{fig:APFComparisonARCH}
as a function of the number of samples $N$.
Interestingly enough, 
our weighted independent resampling algorithm 
which produces $\widehat{\Theta}^{\rm I-SIR-w}_{N,k}$
has the same performances as the FA-APF algorithm 
when $N \geq 15$, 
without using the predictive likelihood
$p(y_k|x_{k-1})$ 
nor the optimal importance distribution
$p(x_k|x_{k-1},y_k)$. 
It means that the mixture pdf $\tilde{q}_N$ 
which has been interpreted in section \ref{paragraph-interpretation}
is indeed as relevant as the target mixture \eqref{filtrage-3};
so in general models where the FA-APF is no longer computable,
one can expect that our estimate $\widehat{\Theta}^{\rm I-SIR-w}_{N,k}$ would
give a performance close to that deduced from FA-APF.
Indeed, one advantage of the mixture pdf $\tilde{q}_N$
deduced from the resampling mechanism is that its interpretation
does not depend on the importance distribution $q_{i,k}$
which has been chosen and that it is possible to sample
from it in general hidden Markov models \eqref{HMC}.
We also observe that re-weighting the final samples
is beneficial w.r.t. attributing uniform weights.
In order to analyze the behavior of the weights 
associated to our estimate 
$\widehat{\Theta}^{\rm I-SIR-w}_{N,k}$,
we compute the 
normalized effective sample size defined as
$N_{norm,eff} = \frac{1}{N\sum_{i=1}^N (w_k^i)^2}$.
In Fig. \ref{fig:CvgPQtildeARCH}, we display 
the time-averaged normalized effective sample size.
It can be observed that $N_{norm,eff}$ tends to $1$
as $N$ increases, meaning that these weights tend to become uniform,
so estimates $\widehat{\Theta}^{\rm I-SIR}_{N,k}$
and $\widehat{\Theta}^{\rm I-SIR-w}_{N,k}$ become close when $N$ is large.


\subsection{Tracking from range-bearing measurements}
\label{simu-polar}
We now study the performance of our algorithms 
in a tracking scenario with range-bearing measurements.
We look for estimating the state vector 
$X_k = [p_{x,k}, \dot{p}_{x,k}, p_{y,k}, \dot{p}_{y,k}]^{T}$ 
(position and velocity  in Cartesian coordinates) of 
a target from noisy range-bearing measurements $y_k$. 
The pdfs in model \eqref{HMC} 
associated with this tracking problem are
$f_{k}(x_k|x_{k-1})=\mathcal{N}(x_k;{\bf F}x_{k-1}; {\bf Q})$
and $g_k(y_k|x_k)=\mathcal{N}(y_k;
\begin{pmatrix}
\sqrt{p^{2}_{x,k} + p^{2}_{y,k}} \\
\arctan{\frac{p_{y,k}}{p_{x,k}}}
\end{pmatrix};{\bf R})$ where $\tau=1$, $\mathbf{R}=
\begin{pmatrix}
\sigma^{2}_{\rho} & 0 \\
0 & \sigma^{2}_{\theta} \\
\end{pmatrix} \text{, }$
\begin{eqnarray*}
\mathbf{F}&=&\begin{pmatrix}
1 & \tau & 0 & 0 \\
0 & 1 & 0 & 0 \\
0 & 0 & 1 & \tau \\
0 & 0 & 0 & 1 
\end{pmatrix} \text{, } 
\mathbf{Q}= \sigma^{2}_{Q}
\begin{pmatrix}
\frac{\tau^{3}}{3} & \frac{\tau^{2}}{2} & 0 & 0 \\
\frac{\tau^{2}}{2} & \tau & 0 & 0 \\
0 & 0 & \frac{\tau^{3}}{3} & \frac{\tau^{2}}{2} \\
0 & 0 & \frac{\tau^{2}}{2} & \tau \\
\end{pmatrix} \text{.} 
\end{eqnarray*}
The conditional importance distribution 
used to sample particles is the transition pdf 
$q(x_k|x_{0:k-1})=f_{k}(x_k|x_{k-1})$;
so the importance weights $w_k^i$ at time $k$
are proportional to $w_{k-1}^ig(y_k|x_k^i)$.
We compute $\widehat{\Theta}^{\rm SIS}_{N,k}$ (see \eqref{theta-sis-seq}),
$\widehat{\Theta}^{\rm I-SIR}_{M,k} $ (see \eqref{theta-isir-seq}),
$\widehat{\Theta}^{\rm I-SIR-w}_{M,k}$ (see \eqref{Theta-ind-2}) 
with $N=\frac{M^2+M}{2}$ to set the number of sampling operations.
We also compare these estimates with $\widehat{\Theta}^{\rm IPF}_{N,k}$
deduced from the Island PF with $5$ islands and $N/5$ particles per island.

\begin{figure}[htbp!]
\centering
\begin{subfigure}[$\sigma_{Q} = \sqrt{10}$, $\sigma_{\rho} = 0.25$ and $\sigma_{\theta} = \frac{\pi}{720}$]
{
\centering
\includegraphics[scale=0.4]{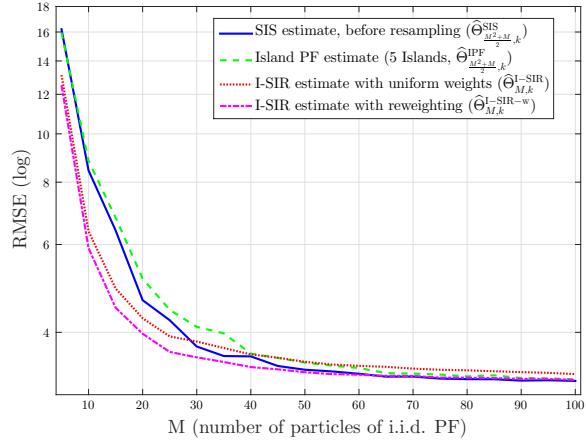}

\label{fig:PolarTargetTrackingDynamic}
}
\end{subfigure}

\begin{subfigure}[$\sigma_{Q} = \sqrt{10}$, $\sigma_{\rho} = 0.05$ and $\sigma_{\theta} = \frac{\pi}{3600}$]
{
\centering
\includegraphics[scale=0.4]{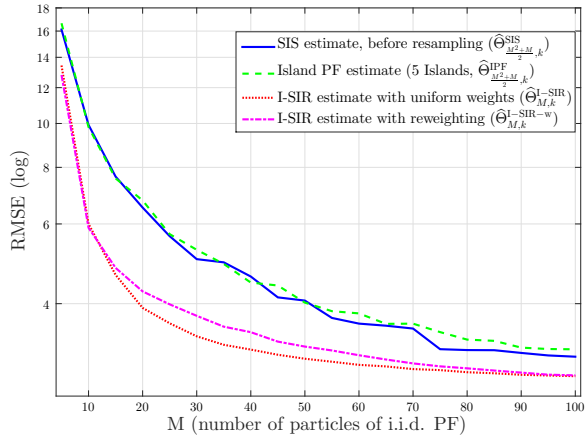}

\label{fig:PolarTargetTrackingDynamic2}
}
\end{subfigure}
\caption{Target tracking model from range-bearing measurements - (a) the independent resampling procedure
with final weighting outperforms the other estimates and is particularly interesting when the number 
of final samples is weak - (b) in the informative case, all estimates suffer from the degeneration of
the importance weights except that based on the unweighted independent resampling algorithm. 
To achieve the same performances as $\widehat{\Theta}^{\rm I-SIR-w}_{M,k}$ with $M=20$, 
the classical PF uses $N=(50^2+50)/2=1275$ samples} 
\end{figure}

The results are displayed for two set of parameters. Fig.
\ref{fig:PolarTargetTrackingDynamic} corresponds to the case where 
$\sigma_{Q}=\sqrt{10}$, $\sigma_{\rho} = 0.25$ and $\sigma_{\theta} = \frac{\pi}{720}$
while Fig \ref{fig:PolarTargetTrackingDynamic2} corresponds to a very informative
case where $\sigma_{Q} = \sqrt{10}$, $\sigma_{\rho} = 0.05$ and $\sigma_{\theta} = \frac{\pi}{3600}$. 
For the first configuration, we observe that 
$\widehat{\Theta}^{\rm I-SIR-w}_{M,k}$ outperforms the other estimates and
improves $\widehat{\Theta}^{\rm I-SIR}_{M,k}$ which does not rely on weighted
samples. Compared to the classical SIS estimate, $\widehat{\Theta}^{\rm I-SIR}_{M,k}$
gives better performance as long as the number of samples 
$M$ is weak ($M<30$, so $N<465$) but is next outperformed
when the number of samples is large. As shown in Fig. \ref{fig:PolarTargetTrackingDynamic2},
when the observations become informative, $\widehat{\Theta}^{\rm I-SIR}_{M,k}$ gives the best
performances. Contrary to $\widehat{\Theta}^{\rm SIS}_{N,k}$ and $\widehat{\Theta}^{\rm IPF}_{N,k}$
our estimate does not suffer from the degeneration of the importance weights. 
Indeed when the measurements are informative (and so the likelihood is sharp), few importance weights have 
a non null value. However, the independent resampling procedure ensures the
diversity of the final samples when we use uniform weights.
Concerning $\widehat{\Theta}^{\rm I-SIR-w}_{M,k}$, remember that it relies
on the MC approximation \eqref{poids-approche}. A close analysis of \eqref{poids-approche}
when the likelihood is sharp shows that the final weights tend to be null except that
of the particle with the larger likelihood; consequently, in this case 
the estimate $\widehat{\Theta}^{\rm I-SIR-w}_{M,k}$ is affected by the
lack of diversity.


\subsection{High dimensional problems}
\label{simu-highdim}
We finally study the impact of the dimension of the hidden state ${X_k}$.
We consider a state vector of dimension $m=4\times l$, $x_k=[p_{x,k}^1, \dot{p}_{x,k}^1, p_{y,k}^1, \dot{p}_{y,k}^1, \cdots,p_{x,k}^{l}, \dot{p}_{x,k}^l, p_{y,k}^l, \dot{p}_{y,k}^l ]^{T}$.
Each component  $x_k^l=[p_{x,k}^l,\dot{p}_{x,k}^l, p_{y,k}^l, \dot{p}_{y,k}^l ]^{T}$ 
evolves independently from all the other components,
according to $f_{k|k-1}(x_k|x_{k-1})=\mathcal{N}(x_k;{\bf F}x_{k-1};{\bf Q})$ where 
$\tau=1$,
\begin{eqnarray*}
\mathbf{F}&=&\begin{pmatrix}
1 & \tau & 0 & 0 \\
0 & 1 & 0 & 0 \\
0 & 0 & 1 & \tau \\
0 & 0 & 0 & 1 
\end{pmatrix} \text{, } 
\mathbf{Q}= \sigma^{2}_{Q}
\begin{pmatrix}
\frac{\tau^{3}}{3} & \frac{\tau^{2}}{2} & 0 & 0 \\
\frac{\tau^{2}}{2} & \tau & 0 & 0 \\
0 & 0 & \frac{\tau^{3}}{3} & \frac{\tau^{2}}{2} \\
0 & 0 & \frac{\tau^{2}}{2} & \tau \\
\end{pmatrix} \text{. } 
\end{eqnarray*}
Each component is observed independently via 
$g_k(y_k|x_k)=\mathcal{N}(y_k;{\bf H}x_k;{\bf R})$ where
\begin{eqnarray*}
\mathbf{H}&=&\begin{pmatrix}
1 & 0 & 0 & 0 \\
0 & 0 & 1 & 0 \\
\end{pmatrix} \text{, } 
\mathbf{R}=
\begin{pmatrix}
\sigma_{x}^2 & 0 \\
0 & \sigma_{y}^2 \\
\end{pmatrix} \text{. } 
\end{eqnarray*}
Again, we compute the estimate
based on classical PF
$\widehat{\Theta}^{\rm SIS}_{N,k}$ (see \eqref{theta-sis-seq}).
It is well known that the PF tends to 
degenerate when the dimension of the hidden state increases.
We also compute $\widehat{\Theta}^{\rm I-SIR}_{M,k} $ (see \eqref{theta-isir-seq}),
$\widehat{\Theta}^{\rm I-SIR-w}_{M,k}$ (see \eqref{Theta-ind-2}) 
with  $N=\frac{M^2+M}{2}$ for $M=100$ and $M=1000$ as a function
of the dimension $m$ to see how the dimension impacts 
our estimate and the classical PF estimate.

\begin{figure}[htbp!]
\centering
\includegraphics[scale=0.47]{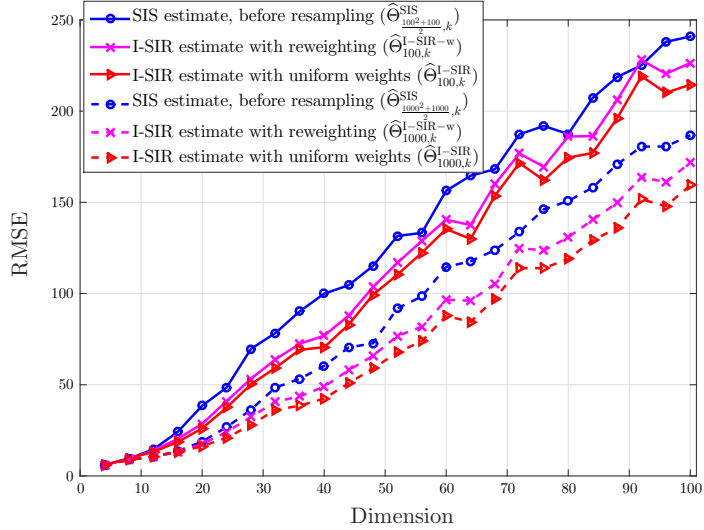}
\caption{Multi-dimensional linear Gaussian model - $\sigma^{2}_{Q} = 25$, $\sigma_{x}^2 = 4$ and $\sigma_{y}^2 = 4$.
The estimates of interest are compared as a function of the dimension $m$ of the hidden state $x_k$ for a fixed number of sampling operations. The independent resampling mechanism limits the impact of the large dimension $m$ and estimate $\widehat{\Theta}^{\rm I-SIR}_{1000,k}$ when $m=46$ has the same performance as $\widehat{\Theta}^{\rm I-SIR}_{500500,k}$ when $m=32$.  }
\label{fig:CartesianTargetTrackingVaryingDim}
\end{figure}

The results are displayed in Fig. \ref{fig:CartesianTargetTrackingVaryingDim}.
It can be seen that the estimates $\widehat{\Theta}^{\rm I-SIR}_{M,k}$ 
and $\widehat{\Theta}^{\rm I-SIR-w}_{M,k}$ outperform $\widehat{\Theta}^{\rm I-SIS}_{M,k}
$ more and more significantly as the dimension increases,
due to the local impoverishment phenomenon.
First, $\widehat{\Theta}^{\rm I-SIR-w}_{M,k}$ outperforms $\widehat{\Theta}^{\rm I-SIR}_{M,k}$
as long as the dimension of the hidden state is low 
($m = 4$ and $m = 8$); when $m$ increases,
the estimate based on weighted samples from $\tilde{q}_N$ limits the 
degeneration phenomenon w.r.t. that based on weighted samples from $q$
but using unweighted samples when the dimension is large ensures
the diversity and gives better performances. 
Note that the dependent and independent SIR algorithms give approximately 
the same performance when $m$ is low but
the gap between the dependent and the independent SIR estimates
increases with the dimension.

%

\section{Conclusion}
SMC algorithms in Hidden Markov models are based on the sequential application of the IS principle.
However the direct sequential application of the IS principle leads to the degeneration
of the weights,
against which multinomial resampling has been proposed.
This rejunevation scheme, which is now routinely used in SIR algorithms,
enables to discard particles (or trajectories) with low weights, 
but particles with large weights will be resampled several times, which leads to dependency and support degeneracy.
In this paper we thus revisited the resampling step used in the classical SIR algorithms.
We first addressed the static case, 
showed that the particles sampled by Rubin's SIR mechanism are dependent samples drawn from some pdf $\tilde{q}_N$,
and proposed an alternative sampling mechanism which produces independent particles drawn from that same marginal pdf $\tilde{q}_N$.
This set of independent samples enables us to build a moment estimator which outperforms the classical SIR-based one, 
both from a non-asymptotical and an asymptotical points of view.
Finally the succession of the sampling, weighting and resampling steps 
indeed transforms an elementary instrumental pdf $q$ into 
a compound importance distribution $\tilde{q}_N = \phi(p, q, N)$, 
which leads us to reweight the (originally unweighted) resampled particles $x^i$
by post-resampling weights proportional to $\frac{p(x^i)}{\tilde{q}(x^i)}$.
Such post-resampling weights cannot be computed exactly, but can easily be estimated 
by recycling the extra MC samples which were needed for producing the independently resampled particles.

We next adapted this methodology to the dynamic case, 
in order to estimate a moment of interest in an hidden Markov model.
The computation of the post-resampling weights is more challenging than in the static case, 
but reinterpreting our independent resampling scheme as the first step of a particular APF algorithm 
enables us to make full use of the APF methodology 
and so to reweight the final samples via the second-stage APF weights. 
Finally we validated our discussions by computer-generated experiments
and carefully took into account the computational budget.
Simulations in model where the FA-APF algorithm is computable 
show that the independent resampling gives a performance
close to the FA-APF algorithm. Consequently, 
it confirms the relevance of the instrumental mixture pdf used implicitly 
by the independent resampling PF which can be used in any hidden Markov model
since it not require to compute the predictive likelihood nor 
the optimal importance distribution. 
Finally independent PF gives very satisfying results when applied in highly informative models 
which are challenging for classical PF and limits the degeneration phenomenon in high dimensional 
models.

\appendix
\section*{Proof of Proposition \ref{prop-q-tilde-static}}
Let $A$ be any Borel set.    
Let 
$\mathds{1}_A(x) = 1$
if $x \in A$ and
$0$ otherwise.
Then for any $l$, $1\leq l \leq M_N$, 
\begin{align*}
& \Pr(X^l \in A)    \\
&= \int_{\RR^N} [\sum_{i=1}^N w^i(\tilde{x}^1, \cdots, \tilde{x}^N) \mathds{1}_{A}(\tilde{x}^i)] \prod_{j=1}^N q(\tilde{x}^j) {\rm d}{\tilde {\bf x}}^{1:N}   \\
& =
\sum_{i=1}^N \int_{\RR^N}  w^i(\tilde{x}^1, \cdots, \tilde{x}^N) \mathds{1}_{A}(\tilde{x}^i) \prod_{j=1}^N q(\tilde{x}^j) {\rm d}{\tilde {\bf x}}^{1:N}  \\
& = \!\! 
\sum_{i=1}^N \! \int_{A}  
[ \int_{\RR^{N-1}} \!\!\!\!\!\!\!\!\! w^i(\tilde{x}^1, \cdots, \tilde{x}^N) \prod_{\underset{j \neq i}{j=1}}^N q(\tilde{x}^j) {\rm d}{\tilde {\bf x}}^{1:i-1,i+1:N} ] q(\tilde{x}^i) {\rm d}{\tilde x}^{i}    \\
& =
\sum_{i=1}^N  \int_{A}  h_N(\tilde{x}^i) q(\tilde{x}^i) {\rm d}{\tilde x}^{i}   \\
& =
\int_{A}  N h_N(\tilde{x}) q(\tilde{x}) {\rm d}{\tilde x} ,
\end{align*}
so $X^l$ has pdf $\tilde{q}_N$ w.r.t. Lebesque measure.

\section*{Proof of Proposition \ref{prop-static-2}}
Let $X^{i}$ (for any $i$, $1 \leq i \leq M_N$
be produced by the classical SIR mechanism. 
Then 
\begin{equation}
\label{E-SIR-SIWR}
{\rm E}(f(X^{i}))|
\tilde{\bf x}^{1:N}
)
=
\widehat{\Theta}^{\rm IS}_{N}.
\end{equation}
So
$
{\rm E}(\widehat{\Theta}^{\rm SIR}_{M_N})|
\tilde{\bf x}^{1:N}
)
=
\widehat{\Theta}^{\rm IS}_{N},
$
and
${\rm E}(\widehat{\Theta}^{\rm SIR}_{M_N}) =$
${\rm E}(\widehat{\Theta}^{\rm IS}_{N})$.
On the other hand 
${\rm E}(\widehat{\Theta}^{\rm I - SIR}_{M_N}) =$
${\rm E}(\widehat{\Theta}^{\rm SIR}_{M_N})$,
whence 
\eqref{egal-esperances}.
Next 
\begin{eqnarray}
\label{var-SIR-somme}
{\rm var}(\widehat{\Theta}^{\rm SIR}_{M_N})
& = &
\frac{1}{M_N^2}
\sum_{i=1}^{M_N}
{\rm var}(f(X^{i}))
+ 
\frac{1}{M_N^2}
\sum_{\stackrel{\scriptstyle k,l=1}{k \neq l}}^{M_N} \!\!
{\rm Cov}(f(X^{k}),f(X^{l})).
\end{eqnarray}
in which $X^{i} \sim \tilde{q}_N$ for all $i$.
The first term is equal to ${\rm var}(\widehat{\Theta}^{\rm I - SIR}_{M_N})$.
Let us compute the second term.
For all $k$, $l$, $1 \leq k, l, \leq M_N$
with
$k\neq l$,
$
{\rm E}(
f(X^{k})f(X^{l})|
\tilde{\bf x}^{1:N}
)
=
(
\widehat{\Theta}^{\rm IS}_{N}
)^2,
$
so 
$
{\rm E}(f(X^{k})f(X^{l})) 
=
{\rm E}(
{\rm E}(
f(X^{k})f(X^{l})|
\tilde{\bf X}^{1:N}
)
)
=
{\rm E}
((\widehat{\Theta}^{\rm IS}_{N}
)^2).
$
Using
\eqref{E-SIR-SIWR} again,
we conclude that 
$
{\rm Cov}(f(X^{k}),f(X^{l}))
=
{\rm var}(\widehat{\Theta}^{\rm IS}_{N}),
$
whence
\eqref{dependent-vs-independent}.

\section*{Proof of Theorem \ref{prop-tcl}}

We first introduce the following notations:
\begin{align}
\Theta(f)&=\int f(x)p(x) {\rm d}x \text{,} \\
\widehat{\Theta}^{\rm IS}_N(f)&= \sum_{i=1}^N \frac{\frac{p(X^i)}{q(X^i)}}{\sum_{j=1}^N \frac{p(X^j)}{q(X^j)}}f(X^i) \text{, } X^i\overset{\rm i.i.d.}{\sim} q(.) \text{,} \\
\widehat{\Theta}^{\rm I-SIR}_{M_N}(f) &= \frac{1}{M_N} \sum_{i=1}^{M_N} f(\overline{X}^i) \text{, } \overline{X}^i \overset{\rm i.i.d.}{\sim} \tilde{q}_N(.) \text{,}  
\end{align}
and we will assume that ${\rm E}(\widehat{\Theta}^{\rm IS}_N(f^2))$ is finite.

Using ${\rm E}(\Theta^{\rm I-SIR}_{M_N}(f))={\rm E}(\Theta^{\rm IS}_N(f))$, 
we have
\begin{align}
\label{equation-proof-0}
& \sqrt{M_N}\left(\widehat{\Theta}^{\rm I-SIR}(f)-\Theta(f)\right) = A_N + B_N \text{,} \\
& A_N = \sqrt{M_N}(\widehat{\Theta}^{\rm I-SIR}(f)-{\rm E}(\widehat{\Theta}^{\rm I-SIR}(f))) \text{,} \\
& B_N = \frac{\sqrt{M_N}}{\sqrt{N}} {\rm E}(\sqrt{N}(\widehat{\Theta}^{\rm IS}_N(f)-\Theta(f))) \text{.}
\end{align}

Our objective is to show that $A_N$ 
converges to a centered Gaussian distribution 
with variance ${\rm var}_p(f(X))$ and that $B_N$ converges to $0$.

\subsection*{Convergence of $B_N$}
We have recalled (see \eqref{LLN-IS}) that under mild assumptions \cite{Gewecke}
\begin{equation*}
\sqrt{N} (\widehat{\Theta}^{\rm IS}_N(f)-\Theta(f))  \overset {\mathcal{D}} {\rightarrow}  \mathcal{N} \left(0,{\rm E}_q \left(\frac{p^2(X)}{q^2(X)}(f(X)-\Theta)^2\right) \right) \text{.}
\end{equation*}
According to Theorem 9.1.10 in \cite{Cappeetal},
${\rm E}(|\sqrt{N} (\widehat{\Theta}^{\rm IS}_N(f)-\Theta(f))|^2)$ is bounded and so its upper bound 
is finite. 
According to the corollary of Theorem 25.12 in \cite{billingsley1995probability}, it is ensured that
$\sqrt{N}{\rm E}((\widehat{\Theta}^{\rm IS}_N(f)-\Theta(f))) \rightarrow 0$; consequently
\begin{equation}
\label{equation-proof-1}
\frac{\sqrt{M_N}}{\sqrt{N}} {\rm E}(\sqrt{N}(\widehat{\Theta}^{\rm IS}_N(f)-\Theta(f))) \rightarrow 0 \text{.}
\end{equation}

\subsection*{Convergence of $A_N$}
$A_N$ reads 
\begin{equation}
\sqrt{M_N} \left( \frac{1}{M_N} \sum_{i=1}^{M_N} f(\overline{X}^i)- {\rm E}(f(\overline{X}^i)) \right) \text{.}
\end{equation}
To prove the convergence when $N \rightarrow \infty$,
we need a CLT for triangular arrays and we use the version presented
in Theorem 9.5.13 of \cite{Cappeetal}. The required assumptions are:
\begin{enumerate}
\item $\{\overline{X}^i\}_{i=1}^{M_N}$ are independent;
\item $\frac{1}{M_N} \sum_{i=1}^{M_N} {\rm E}(f^2(\overline{X}^i))- ({\rm E}(f(\overline{X}^i)))^2 \rightarrow {\rm var}_p(f(X))$;
\item for any positive $C$, $\frac{1}{M_N} \sum_{i=1}^{M_N} {\rm E}(f^2(\overline{X}^i)\mathds{1}_{|f(\overline{X}^i)| \geq C}) \rightarrow \Theta(f^2\mathds{1}_{|f| \geq C})$.
\end{enumerate}

Assumption 1) is satisfied since $\{\overline{X}^i\}_{i=1}^{M_N}$ are i.i.d. from $\tilde{q}_N$. 
Next, ${\rm E}(f(\overline{X}^i))={\rm E}_{\tilde{q}_N}(f(\overline{X}))$ which coincides with 
${\rm E}(\widehat{\Theta}^{\rm IS}_N(f))$.
Using again Theorem 9.1.10 of \cite{Cappeetal} and Theorem 25.12 of \cite{billingsley1995probability},
${\rm E}(\widehat{\Theta}^{\rm IS}_N(f)) \rightarrow \Theta(f)$ when $N \rightarrow \infty$.
With the same argument, ${\rm E}(f^2(\overline{X}^i)) \rightarrow \Theta(f^2)$. 
Consequently, assumption 2) is satisfied since
\begin{align*}
& \frac{1}{M_N} \sum_{i=1}^{M_N} {\rm E}(f^2(\overline{X}^i))- ({\rm E}(f(\overline{X}^i)))^2 = 
{\rm E}_{\tilde{q}_N}(f^2(\overline{X}))- ({\rm E}_{\tilde{q}_N}(f(\overline{X})))^2 \rightarrow \Theta(f^2) - (\Theta(f))^2=
{\rm var}_p(f(X)).
\end{align*} 
Finally, ${\rm E}(f^2(\overline{X}^i)\mathds{1}_{|f(\overline{X}^i)| \geq C})={\rm E}(\widehat{\Theta}^{\rm IS}_N(f^2\mathds{1}_{|f| \geq C}))$ which converges to $ \Theta(f^2\mathds{1}_{|f| \geq C})$ and assumption 3) is satisfied.
Consequently,
\begin{equation}
\label{equation-proof-2}
\sqrt{M_N}(\frac{1}{M_N} \sum_{i=1}^{M_N} f(\overline{X}^i)- {\rm E} (f(\overline{X}^i))) \overset{\mathcal{D}}{\rightarrow} \mathcal{N}(0,{\rm var}_p(f(X))) \text{.}
\end{equation}

Combining \eqref{equation-proof-1}, \eqref{equation-proof-2} and \eqref{equation-proof-0} we obtain \eqref{TCL-1}.

%
%

\ifCLASSOPTIONcaptionsoff
  \newpage
\fi

%
%

\bibliographystyle{IEEEtran}
\bibliography{yohan}

%

%
%
%

\end{document}